\documentclass{aa}  

\usepackage{graphicx}
\usepackage{lscape}
\usepackage{amsmath}
\usepackage{txfonts}
\usepackage{url}
\usepackage{natbib}
\usepackage[usenames,dvipsnames,svgnames,table]{xcolor}
\usepackage{ulem}

\newcommand{\kms}{km~s$^{-1}$}

\begin{document} 

\title{Spectrally resolved detection of sodium in the atmosphere of\\
HD\,189733b with the HARPS spectrograph\thanks{using observations with the Harps spectrograph from the ESO 3.6m installed at La Silla, Chile, under the allocated programmes 072.C-0488(E), 079.C-0828(A) and 079.C-0127(A).}}
\author{A. Wyttenbach, D. Ehrenreich, C. Lovis, S. Udry, F. Pepe}
\authorrunning{Wyttenbach et al.}
\titlerunning{Detection of sodium in the atmosphere of HD\,189733b with HARPS}
\institute{Geneva Observatory, University of Geneva, ch. des Maillettes 51, CH-1290 Versoix, Switzerland\label{inst1}\\
\email{aurelien.wyttenbach@unige.ch}}

\date{Received 2015-01-23; accepted 2015-03-14} 
\abstract
   {Atmospheric properties of exoplanets can be constrained with transit spectroscopy. At low spectral resolution, this technique is limited by the presence of clouds. The signature of atomic sodium (\ion{Na}{i}), known to be present above the clouds, is a powerful probe of the upper atmosphere, where it can be best detected and characterized at high spectral resolution.}
   {Our goal is to obtain a high-resolution transit spectrum of HD~189733b in the region around the resonance doublet of \ion{Na}{i} at 589~nm, to characterize the absorption signature previously detected from space at low resolution.}
   {We analyze archival transit data of HD~189733b obtained with the HARPS spectrograph ($\mathcal{R}=115\,000$) at ESO 3.6-meter telescope. We perform differential spectroscopy to retrieve the transit spectrum and light curve of the planet, implementing corrections for telluric contamination and planetary orbital motion. We compare our results to synthetic transit spectra calculated from isothermal models of the planetary atmosphere.}
   {We spectrally resolve the \ion{Na}{i} D doublet and measure line contrasts of $0.64\pm0.07\%$ (D2) and $0.40\pm0.07\%$ (D1) and FWHMs of $0.52\pm0.08~\AA$. This corresponds to a detection at the 10-$\sigma$ level of excess of absorption of $0.32\pm0.03\%$ in a passband of $2\times0.75\ \AA$ centered on each line. We derive temperatures of $2\,600\pm600$~K and $3270\pm330$~K at altitudes of $9\,800\pm2\,800$ and $12\,700\pm2\,600$~km in the \ion{Na}{i} D1 and D2 line cores, respectively. We measure a temperature gradient of $\sim0.2$~K~km$^{-1}$ in the region where the sodium absorption dominates over the haze-absorption, from comparison with theoretical models. We also detect a blueshift of $0.16\pm0.04\ \AA$ (4~$\sigma$) in the line positions. This blueshift may be due to winds blowing at $8\pm2$~\kms\ in the upper layers of the atmosphere.}
   {We demonstrate the relevance of studying exoplanet atmospheres with high-resolution spectrographs mounted on 4-meter-class telescopes. Our results pave the way towards in-depth characterization of physical conditions in the atmospheres of many exoplanetary systems with future spectrographs such as ESPRESSO on the VLT or HiReS and METIS on the E-ELT.}


\maketitle

\section{Introduction}

After two decades and over a thousand exoplanet detections, we have entered an era of characterization of these remote worlds. More and more exoplanets have precise mass and radius measurements thanks to simultaneous radial-velocity and transit observations, providing us with estimates of mean densities and constraints on possible bulk compositions. Meanwhile, detections and studies of exoplanet atmospheres represent the only direct observable window we have on the physical and chemical properties of these planets. Characterization of exoplanets and comparative planetology are now progressing well by virtue of cutting-edge instrumentation. Among these observations, spectra of exoplanets are among the best products for in-depth comprehension of atmospheric and surface conditions.

Shortly after the first indirect exoplanet discoveries, early observational campaigns encouraged by theoretical work \citep{Marley1999,Seager2000,Brown2001}, attempted to detect reflected light from the dayside \citep{Charbonneau1999} or absorption through limb transmission \citep{Moutou2001}. Observations of HD\,209458, the unique transiting system known at that time, with the Space Telescope Imaging Spectrograph (STIS) on board the 2.4-m Hubble Space Telescope (HST), allowed \citet{Charbonneau2002} to detect atmospheric sodium (\ion{Na}{i}) at 589~nm. This measurement was performed with a spectral resolution of $\mathcal{R}\equiv\lambda/\Delta\lambda\sim5\,500$ (or $\sim55$~\kms). Extra absorption of 0.023$\pm0.006\%$ and 0.013$\pm0.004\%$ have been measured during transits over spectral bins of 12~\AA\ and 38~\AA, respectively. The \ion{Na}{i} doublet lines D1 ($\lambda589.5924$~nm) and D2 ($\lambda588.9951$~nm) are not resolved in these bins.

In addition to important detections of other atomic and molecular signatures \citep[\textit{e.g.}][]{Vidal2003,Deming2013}, the sodium signature in HD\,209458b has remained one of the most robust examples of atmospheric characterization \citep{Madhusudhan2014,Heng2014,Pepe2014}. It has been confirmed by independent analysis of the same data set \citep{Sing2008a,Sing2008b}. In the early days, ground-based observations of transiting systems with high-resolution spectrographs mounted on 8-meter-class telescope were also quickly tested in the perspective of confirming detections and resolving transmission spectra, unfortunately without any positive results \citep[\textit{e.g.}][]{Narita2005}. Indeed, even with high-enough signal, no good observational strategy or data analysis were found to overcome terrestrial atmospheric variation and telluric lines contamination that transmission spectra undergo. Later, careful data (re-)analysis made by \citet{Redfield2008} and \citet{Snellen2008} enabled to beat down systematics due to Earth atmosphere and led to the first ground-based detection of exoplanet atmospheric signature. Indeed the \ion{Na}{i} doublet was detected for the first time in HD\,189733b with the High Resolution Spectrograph (HRS; $\mathcal{R}\sim60\,000$) mounted on the 9-m Hobby-Eberly Telescope and confirmed in HD\,209458b using the High Dispersion Spectrograph (HDS; $\mathcal{R}\sim45\,000$) on the 8-m Subaru telescope. Furthermore, the analysis of the space-based detection of the sodium signature of these two planets by \citet{Sing2008b}, \citet{Vidal2011}, and \citet{Huitson2012}, demonstrated the potential to resolve the transmitted star light by the planet atmosphere. Despite these encouraging works, the presence of scattering hazes and clouds \citep{Lecavelier2008, Lecavelier2008b, Pont2013} in several exoplanets prevented the detection of major chemical constituents at low-to mid-resolution even from space \citep[\textit{e.g.}][]{Kreidberg2014, Sing2015}. The whole potential of high-resolution spectroscopy with $R\sim100\ 000$ applied to transiting and non-transiting system was shown when $\mathrm{H_2O}$ and CO molecular bands were detected on account of cross-correlation on hundreds of resolved individual transitions \citep{Snellen2010,Snellen2014,Brogi2012,Birkby2013}. This state-of-the-art method brought back attention to the importance of transit spectroscopy observations from the ground.

Transit spectroscopy with ground-based telescopes is often used to gauge the Rossiter-McLaughlin effect during an exoplanet transit. This is accomplished with high-precision velocimeters. As data are acquired during transit, we can think about measurements of exoplanet atmosphere properties with the same data. However, observations with simultaneous-calibrated or self-calibrated velocimeter are not equally valuable because of the use of an iodine cell. The absorption of the light by the iodine cell act as a contaminant. The iodine lines prevent measurements of several spectral features especially around the \ion{Na}{i} lines. Thus the use of stabilized spectrographs with simultaneous calibration makes possible to measure the Rossiter-McLaughlin effect and the transmission spectrum of an exoplanet at the same time.

In this paper, we will describe new results obtained with the HARPS spectrograph by analyzing existing high-quality Rossiter-McLaughlin observations of HD\,189733b. For the first time, features in an exoplanet atmosphere are resolved with a 4-m-class telescope. In Sect. \ref{Sec_Obs}, we will present the observations, the data reduction, the telluric correction and the method to measure exo-atmospheric signals. In Sect. \ref{Sec_Results}, we will describe the results obtained on the measure of the \ion{Na}{i} D absorption excess in HD\,189733b. In Sect. \ref{Sec_Discuss}, we will discuss the implication of our detection, eventually we will compare our results with theory. 

\section{Observations, data reduction and methods}\label{Sec_Obs}

\subsection{HARPS observations}
HD\,189733 \citep{Bouchy2005} was observed with the HARPS echelle spectrograph (High-Accuracy Radial-velocity Planet Searcher) on the ESO 3.6~m telescope, La Silla, Chile. Data were retrieved from the ESO archive from programs 072.C-0488(E), 079.C-0828(A) (PI:Mayor) and 079.C-0127(A) (PI: Lecavelier des Etangs). The data sets are described in Table~\ref{table1} (nights 1,3,4 were employed by \citet{Triaud2009} to measure the Rossiter-McLaughlin effect). In total, four sequences of data cover at least partially the transit of the planet. Among those four sequences, two were obtained using low cadence (600 s exposures), the two others using high cadence (300 s exposures). One low-cadence sequence was affected by bad weather and missed the second half of the transit. For this reason, we decided not to take this sequence (2006-07-29) into account. Our analysis is based on the remaining three data sequences totalizing 99 spectra and 3 planetary transits. Based on the transit ephemeris of \citet{Agol2010} (see Table~\ref{table2}), we identified 46 spectra fully obtained “in-transit” (i.e. fully between the first and fourth contacts). The 53 remaining spectra constitute our “out-of-transit” sample.

\begin{table*}
\caption{Observations log for HD\,189733. Three transit events are analyzed in our work.}
\begin{center}
\begin{tabular}{lccccccccc}
\hline
\rule[0mm]{0mm}{5mm}	& Date & $\#$ Spectra$^1$ & Exp. Time [s] & Airmass & Seeing & SNR$^2$ & SNR$^3$ & Program & Analysis \\
\hline
$\mathrm{Night\ 1}$ & 2006-09-07 & 20 (9/11) & 900 to 600 & 1.6-2.1 & 0.7-1.0 & $\sim170$ & $\sim32$ & 072.C-0488(E) & Yes\rule[0mm]{0mm}{3mm}\\
$\mathrm{Night\ 2}$ & 2007-07-19 & 39 (18/21) & 300 & 2.4-1.6 & 0.6-0.8 & $\sim110$ & $\sim20$ & 079.C-0828(A) & Yes \\
$\mathrm{Night\ 3}$ & 2007-08-28 & 40 (19/21) & 300 & 2.2-1.6 & 0.7-2.0 & $\sim100$ & $\sim18$ & 079.C-0127(A) & Yes \\
$\mathrm{Night\ 4}$ & 2006-07-29 & 12 (5/7) & 600 & 1.8-1.6 & 0.9-1.2 & $\sim140$ & $\sim27$ & 072.C-0488(E) & No \\
\hline
\end{tabular}
\tablefoot{$^1$In parenthesis: the number of spectra taken during the transit (forming our “in-transit” sample) and outside the transit (forming our “out-of-transit” sample). $^2$The signal-to-noise ratio (SNR) per extracted pixel in the continuum near 590 nm. $^3$The SNR in the lines core of the \ion{Na}{i} D doublet (this show the first difficulty of sodium detection in exoplanet atmospheres).}
\end{center}
\label{table1}
\end{table*}

\subsection{Data reduction}
The HARPS observations were reduced with the version 3.5 of the HARPS Data Reduction Software. Spectra are extracted order by order and flat-fielded with the daily calibration set. Each spectral order is blaze-corrected and wavelength-calibrated. All spectral orders from a given two-dimensional echelle spectrum are merged and resampled with a 0.01~$\AA$ wavelength step into a one-dimensional spectrum. Flux is conserved during this step. A reduced HARPS spectrum covers the region between 380~nm and 690~nm with a spectral resolution of $\mathcal{R}\sim115\,000$ or 2.7~\kms. All spectra are referred to the Solar System barycenter rest frame and wavelengths are given in the air.

As we will see in Sect.~\ref{subSec_TransSpec}, the next step is to co-add the in-transit spectra on the one hand, and the out-of-transit spectra on the other hand to build the master-in and master-out spectra. The ratio of the master-in and the master-out yields the transmission spectrum. Then, any changes in the stellar line shape or position during the observations can create spurious signals in the transmission spectra, which can become false-positive of atmospheric signals (i.e., with similar amplitude). This is the case for changes in the line spread function (LSF) of the instrument. This is where the spectrograph design is important: HARPS is a fiber-fed spectrograph stabilized in temperature and pressure, ensuring that changes in the point spread function (PSF) between two consecutive observations lead to negligible changes in the LSF at time scales of one night. Nonetheless, change in the line positions are due to the radial-velocity variation of the star during the transit of the planet ($\sim 50$~m~s$^{-1}$). This must be removed before computing the transmission spectra. Actually, omitting this step for spectral shifts of 0.01~\AA\ or larger creates artificial signals. We simply correct from this effect by shifting the spectra to the null stellar radial velocity (i.e. the mid-transit stellar radial velocity) according to a Keplerian model of the targeted planetary system (see the model parameters in Table~\ref{table2}). Another source of change in the stellar line shape is the Rossiter-McLaughlin effect. As the transit occurs, the planet occults different parts of the stellar disk which have different intrinsic line shapes and shifts. False-positive features will then appear in the transmission spectrum even if the stellar spectra are well aligned. A way to overcome this difficulty is to register spectra uniformly during the transit, so that this effect will be averaged out during a full transit.

\begin{table*}
\caption{Adopted values for the orbital and physical parameters of HD\,189733b.}
\begin{center}
\begin{tabular}{lcclc}
\hline
\rule[0mm]{0mm}{5mm}Parameters & Symbol	&	Values & Units	&References	\rule[0mm]{0mm}{3mm}\\
\hline
Transit epoch (BJD)	&	$T_0$	&	$2454279.436714\pm0.000015$	&	days	&	\citet{Agol2010}\\
Orbital Period	&	$P$	&	$2.21857567\pm0.00000015$	&	days	&	\citet{Agol2010}\\
Planet/star area ratio		&	$(R_p/R_s)^2$	&	$0.02391\pm0.00007$&	&	\citet{Torres2008}\\
Transit duration		&	$t_T$	&	$0.07527\pm0.00037$	&	days	&	\citet{Triaud2009}\\
Impact parameter	&	$b$	&	$0.6631\pm0.0023$	&	$R_{\star}$	&	\citet{Agol2010}\\
Orbital semi-major axis	&	$a$	&	$0.0312\pm0.00037$	&	AU	&	\citet{Triaud2009}\\
Orbital inclination	&	$i$	&	$85.710\pm0.024$	&	degrees	&	\citet{Agol2010}\\
Orbital eccentricity	&	$e$	&	0	&	&	fixed\\
Longitude of periastron	&	$\omega$	&	90	&	degrees	&	fixed\\
Stellar velocity semi-amplitude	&	$K_1$	&	$200.56\pm0.88$	&	$km\ s^{-1}$	&	\citet{Boisse2009}\\
Systemic velocity	&	$\gamma$	&	$-2.2765\pm0.0017$	&	$km\ s^{-1}$	&	\citet{Boisse2009}\\
Stellar mass	&	$M_{\star}$	&	$0.823\pm0.029$	&	$M_{\odot}$	&	\citet{Triaud2009}\\
Stellar Radius	&	$R_{\star}$	&	$0.756\pm0.018$&	$R_{\odot}$	&	\citet{Torres2008}\\
Planet Mass	&	$M_P$	&	$1.138\pm0.025$	&	$M_J$	&	\citet{Triaud2009}\\
Planet Radius	&	$R_P$	&	$1.138\pm0.027$	&	$R_J$	&	\citet{Torres2008}\\
\hline
\end{tabular}
\end{center}
\label{table2}
\end{table*}

\subsection{Telluric correction}\label{TellCorr}
High-resolution spectra recorded from the ground bear the imprint of Earth's atmosphere. In the visible domain covered with HARPS, water vapor and molecular oxygen are the main contributors to this time-variable telluric contamination. Indeed, variation in the transmission of Earth’s atmosphere during a night depends on the airmass and on water column variations in the air. In addition to the contaminating water lines around the \ion{Na}{i} doublet, we expect the observed spectra to also possess a signature of the telluric sodium \citep{Vidal2010}. Even if the telluric sodium undergoes seasonal variations and possibly does not follow the water absorption levels, we make the assumption that, within a night, this telluric sodium absorption behave and can be corrected as other telluric features \citep{Snellen2008,Zhou2012}. Furthermore, visual inspection of our telluric spectra did not reveal any obvious telluric sodium features.

\begin{figure}
\centering
\includegraphics[width=0.47\textwidth]{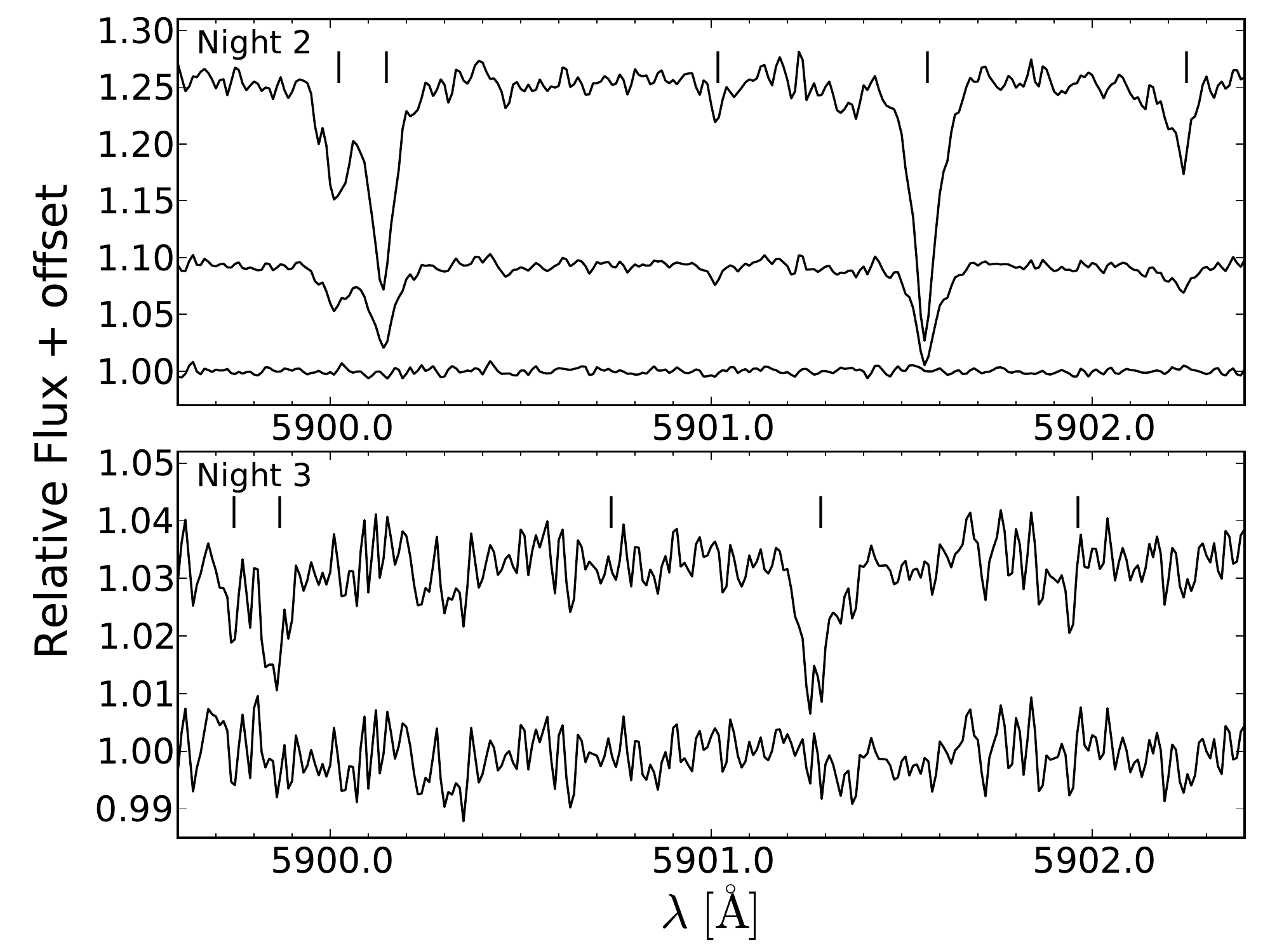}
\caption{Example of the effect of the Earth atmosphere on exoplanetary transmission spectra. This shows the importance of a precise telluric correction. Upper panel: Calculated telluric spectrum $T(\lambda)$ (top), transmission spectrum without any correction (middle), and transmission spectrum with the telluric correction (bottom) for night 2. The telluric lines are dominating all features in the uncorrected transmission spectrum. Lower panel: Transmission spectra with telluric corrections already performed for night 3. Once we apply the telluric correction described in Sect.~\ref{TellCorr} (airmass effect) we get a transmission spectra with weak telluric lines residuals (top). Since residuals of telluric lines are still visible, we apply a second telluric correction (correction of features due to water column variation, see Sect.~\ref{subSec_TransSpec}). This affords us to get a transmission spectrum corrected from the effect of Earth atmosphere (bottom). Note the different vertical scales. Some evident telluric lines are emphasized with vertical lines. The telluric lines are shifted from one night to the other because the spectra are referred to the Solar System barycenter rest frame and were not observed at the same barycentric Earth radial velocity.}
\label{TellCorrFig}
\end{figure}

Telluric contamination produce ubiquitous features in transmission spectra. However, the telluric features can be removed by subtracting high-quality telluric spectra obtained on the same nights. As our observed spectra are referred to the Solar System barycentric rest frame, the telluric lines are shifted during the three different nights by -0.26~\AA, 0.09~\AA\ and -0.19~\AA\ compared to their rest frame. Methods described in \citet{Vidal2010} and \citet{Astudillo-Defru2013} permit us to build high-quality telluric spectra night by night using the HARPS observations themselves. Both methods consider that the variation of telluric lines follows linearly the airmass variation. This is a consequence of the usual hypothesis of a radiative transfer in a plane parallel atmosphere. It implies that apart from different constructions of telluric spectrum these methods yields equivalent telluric spectra. Similarly, we build for each night a reference telluric spectrum $T(\lambda)$ such as
\begin{equation}
T(\lambda) = T^1(\lambda) \equiv e^{Nk_{\lambda}}\
\end{equation}
corresponding to a zenithal atmospheric transmission (airmass $\sec z = 1$). The zenithal optical depth $Nk_{\lambda}$ is derived at each wavelength $\lambda$ by linear regression on the logarithm of the normalized flux $\log F(\lambda)$ as a function of $\sec z$ \citep{Astudillo-Defru2013}. The computed telluric spectra (see Fig.~\ref{TellCorrFig}) are consistent with the atmospheric transmission obtained by \citet{Hinkle2003}.
Note that the telluric spectra are built with the out-of-transit sample only. This is important in order not to over-correct for the possible exoplanet atmosphere we are looking for. However, for night 2, we had to take into account the in-transit sample because the base line was insufficient to compute a high-quality telluric spectrum (during this night, there was no observation before transit).

Taking advantage of the scaling relation between the telluric line strength and the airmass, we can correct spectra for telluric contamination. Corrections are made by rescaling all spectra as if they had been observed at the same airmass. To obtain this spectrum $F_{ref}$ at airmass $\sec z_{ref}$, we simply divided the observed spectra $O_{\sec z}(\lambda)$ taken at airmass $\sec z$ by the telluric reference spectrum to the power $\sec z-\sec z_{ref}$:
\begin{equation}
F_{ref}(\lambda) = O_{\sec z}(\lambda)/T(\lambda)^{\sec z-\sec z_{ref}}\ .
\end{equation}
We fix the reference airmass to the average airmass of the in-transit spectra. This ensures both a telluric correction and minimal flux changes in the reduced spectra. If the quality and the stability of the observation nights are good enough which typically depends on a constant water atmospheric content, this method is sufficient to correct down to the statistical noise level. However, the quality of the nights are generally not sufficient as second-order variations of telluric lines are present and telluric residual can be still visible in transmission spectra (see Fig.~\ref{TellCorrFig} and Sect.~\ref{subSec_TransSpec}). Indeed these observation were not planned to observe exoplanetary atmospheres but Rossiter-McLaughlin effects. This is less demanding in terms of sky quality and requires shorter time baseline. Later in our analysis, we will perform a second telluric correction to remove possible telluric residuals.

\subsection{Transmission spectrum}\label{subSec_TransSpec}
Transit spectroscopy is a differential technique requiring the acquisition of spectra during and outside transit events. Ground-based observations out-of-transit ($O_{out}$) contain the star light absorbed by the terrestrial atmosphere. Observations during transit ($O_{in}$) additionally contain the exoplanet atmosphere transmission diluted into the stellar flux. Stacking telluric-corrected spectra in-transit ($F_{in}$) and out-of-transit ($F_{out}$) allows to obtain master-in ($\mathcal{F}_{in}$) and master-out spectra ($\mathcal{F}_{out}$), respectively. Classical methods obtain transmission spectra ($\mathfrak{R}^{'}=\mathfrak{R}-1=\mathcal{F}_{in}/\mathcal{F}_{out}-1$) of the exoplanet atmosphere by dividing night by night the master spectrum in-transit by the master spectrum out-of-transit \citep{Redfield2008}. We can then analyze exoplanet atmospheres features along wavelengths of interest. Such methods do not consider changes in radial velocity of the \textit{planet}. 

Here, we present a modified approach with respect to \citet{Redfield2008} to correct for this effect. As the planet transits its star, its radial velocity changes typically from $-15$~\kms\ to $+15$~\kms. Therefore, for the sodium the planetary atmospheric absorption lines move from the blue-shifted to the red-shifted part of the stellar lines. In the optical, near the sodium doublet this shift is $\lesssim0.5\ \AA$ during the full transit. We correct this effect in our analysis by shifting the planetary signal to the null radial velocity in the \textit{planet} rest frame (i.e. the radial velocity of the planet in the middle of the transit). Instead of taking the ratio of master spectra, we divide each spectrum in-transit by the master-out, then shift the residual with the computed planetary radial velocity and sum up all the individual transmission spectra. We normalized this sum to unity by a linear fit to the continuum outside the regions of interests. This allows us to compute the transmission spectra as
\begin{equation}
\mathfrak{R}^{'}(\lambda) = \frac{\mathcal{F}_{in}}{\mathcal{F}_{out}} -1 = \sum\limits_{in} \frac{F_{in}(\lambda)}{\sum\limits_{out} F_{out}(\lambda)}\Big|_{Planet\,RV\,shift} - 1\ .
\end{equation}
With this correction, we expect to see the exoplanetary signal at wavelengths corresponding to the systemic velocity rest frame (and not in the solar system rest frame).

A transmission spectrum (with or without radial-velocity correction) corrected from the airmass effect (Sect.~\ref{TellCorr}) can still present some telluric line residuals. This is due to water column variations in the air above the telescope. Then, we applied a second telluric correction to correct from telluric features in the transmission spectra. Because both the transmission spectra and the telluric spectra contain telluric features at corresponding wavelengths, these two spectra are correlated. To eliminate these correlations and the residuals of telluric lines in the transmission spectrum, a linear fit is made between all corresponding fluxes in wavelength of the transmission spectrum and the telluric reference spectrum. The transmission spectrum is then divided by the fit solution. As only telluric residuals are correlated with the telluric spectrum, doing this iteratively eliminates telluric residuals in the transmission spectra without affecting the other part of the spectra \citep{Snellen2008, Snellen2010,Khalafinejad2013}. Practically, one iteration is sufficient to tame atmospheric pollution down to the photon-noise level.

\subsection{Binned atmospheric absorption depth}\label{subSec_BinTS}
In order to compare our results with previous detections of sodium in HD\,189733b \citep{Redfield2008,Jensen2011,Huitson2012} or in other exoplanets \citep{Charbonneau2002,Snellen2008,Sing2008a,Langland2009,Wood2011,Sing2012,Zhou2012,Murgas2014,Nikolov2014,Burton2015}, we calculate the relative absorption depth across various bins in wavelength. Taking into account the systemic velocity, we average the flux around the central passbands ($C$) centered on both lines of the sodium doublet. Here and later in our analysis, all the averages are weighted by the estimated errors on the reduced spectra, which are taken to be random photon noise obeying Poisson statistics. We compare the integrated flux in the transmission spectrum $\mathfrak{R}^{'}(C)$ to bins of similar band widths taken in the transmission spectrum continuum. \citet{Snellen2008} chose adjacent control passbands on the blue ($B$) and red ($R$) side of the central passband. As at high-resolution exoplanet atmospheric lines are likely to be resolved, we prefer to choose absolute reference passbands outside the sodium doublet, but still on the two side (B and R) of the transition \citep[see \textit{e.g.}][]{Charbonneau2002}. Relative depths due to exo-atmospheric absorption are then obtained by the difference of fluxes between the central and the reference passband,
\begin{equation}
\delta(\Delta\lambda) = \overline{\mathfrak{R}^{'}(C)} - \frac{\overline{\mathfrak{R}^{'}(B)} + \overline{\mathfrak{R}^{'}(R)}}{2}\ .
\end{equation}
As small passbands encompass only one line, we average the absorption depth of the two \ion{Na}{i} D lines. When the central passband includes the two lines, we adjusted it on the center of the doublet.

\subsection{Transmission light curve}\label{light_curve_method}
This method consists in directly deriving the relative transmission light curve as a function of time \citep{Charbonneau2002,Snellen2008}. Thus, the absorption excess due to the exoplanet atmospheric limb can be seen as a relative flux decrease during the transit. It differs from the method presented in Sect.~\ref{subSec_TransSpec}. The spectrophotometry is performed on individual spectra for a given spectral bin. We obtain a time-dependent information but the wavelength dependence is lost.

For each telluric-corrected stellar spectrum ($F$), we derive the relative flux at the \ion{Na}{i} D lines by comparing fluxes inside passbands in the center, red and blue part of each lines (see Sect.~\ref{subSec_BinTS} and \ref{Sec_SodiumD} for the description of the passbands):
\begin{equation}
\mathcal{F}_{rel}(t,\Delta\lambda)= \frac{2\times\overline{F(C)}}{\overline{F(B)} + \overline{F(R)}}\ .
\end{equation}
We then normalized the relative time serie to unity. For the smaller passbands, we average, spectrum by spectrum, the relative fluxes of the two \ion{Na}{i} D lines.

Because we always compare together same parts of each spectrum, the relative flux should be constant with time except if absorption by the planetary atmosphere is present during the transit. With this method, we cannot apply any planetary radial-velocity correction. Taking into account the radial-velocity effect would change the central passband, thus it would change the relative flux (not because there is an absorption, but because we do not have a constant central passband). However this is potentially an issue only for the smallest passbands for which we can expect a loss of absorption.

Another type of telluric correction is applied to the relative flux sequence. As the telluric absorption changes with airmass, it is expected to see correlations between relative fluxes and the airmass, especially when the passbands contain telluric lines. The first correction applied on individual stellar spectra (Sect.~\ref{TellCorr}) should have already mitigated this effect. However, some light curve still show correlations with airmass. To remove any residual airmass effect, we model the flux variation as a linear function of the airmass and remove the linear trend. If possible, we only consider out-of-transit data to perform our fit.

\citet{Astudillo-Defru2013} discussed whether or not fitting a transit model to the data can change the measurement of the absorption depth. They showed that the absorption depth is well measured by simple average differences. Indeed, differential stellar limb-darkening do not affect significantly measurement in such narrow passbands. The transmitted flux or relative depth is then given by the difference of the average of the relative flux in and out-of-transit:
\begin{equation}
\delta(\Delta\lambda)= \frac{\quad\overline{\mathcal{F}_{rel}(t_{In}})\quad}{\overline{\mathcal{F}_{rel}(t_{Out})}} - 1\ .
\end{equation}

Integrating the transit light curve over the in-transit duration makes it possible to compare the result with the binned atmospheric absorption depths described in Sect.~\ref{subSec_BinTS}. Both approaches give similar results, because of the normalization processes and the use of photon noise weighted averages in the two respective methods.

\subsection{Systematic effects}\label{Stat_methods}
To measure the uncertainty on the relative absorption depth $\delta(\Delta\lambda)$, we propagated the errors of the reduced spectra through our analysis. We estimated the errors on the reduced spectra as random photon noise obeying Poisson statistics. Systematic effects, on the other hand, are likely to contribute to the total noise budget. We have used different statistical methods, such as empirical Monte-Carlo \citep{Redfield2008} or bootstrapping, to estimate the impact of correlated noise. The basic principle of these methods is to randomize data, to artificially create new set of observations, and to feed them to our data analysis.

Firstly, we followed the empirical Monte-Carlo technic presented in \citet{Redfield2008}. The idea is to see whether or not the measurement is an artifact of the data or if it is really created by the transit and hence due to the planetary companion. We select a sub-sample of spectra of all the spectra available in a night. The choice of this sub-sample is then fixed (we call it a scenario). Next, this new sample is randomly divided in two parts to create an “in-transit“ and an “out-of-transit” simulated data sets. These two data sets always contain the same number of spectra.

As in \citet{Redfield2008}, we explored three scenarios per night. In the first scenario (“out-out”), the spectra are only selected among the nominal out-of-transit spectra. The amount of spectra in the simulated in-transit data compared to the number of spectra in the simulated out-of-transit data are in the same proportion as in the nominal observations. As we choose randomly only among the out-of-transit spectra, we expect to find the distribution of computed relative absorption depth centered at zero. Similarly, an “in-in” scenario is created, but the random in-transit and out-of-transit data are chosen only in the spectrum observed during the transit. As the planetary signal should be present inside every in-transit observation, the final “in-in” distribution is also expected to be centered at zero. Finally an “in-out” scenario is executed. All the out-of-transit observation form a fixed master-out (actually the same than the nominal master-out). We choose randomly a sub-sample of in-transit spectra. Here, we did not fix the number of spectra of the in-transit sub-sample. This later one is composed by half to the totality of the number of nominal in-transit spectra.

How to interpret the computed distribution? \citet{Redfield2008} considered the standard deviation of the “out-out” distribution as the global error on the transmitted signal. Nonetheless, \citet{Astudillo-Defru2013} considered instead the standard deviation of the “in-out” distribution as the error on the transmitted signal. Here, we follow \citet{Redfield2008} and take the “out-out” scenario to infer the error on our measured absorption depth, because it is independent of any planetary signal. Nevertheless, the errors are overestimated with the standard deviation of the “out-out” scenario, because in each iteration only a fraction of the data is used. Thus, to get correct errors on our absorption depth, we divide the standard deviation of the “out-out” scenario by the square root of the ratio of the total number of spectra to the number of spectra out-of-transit.

Secondly, we calculate false alarm probabilities. We randomize spectra and recreate transmission spectra and light curves (i.e. by randomizing the time sequence), and quantify how many randomized data sets yield a signal as significant as the real measurement. This is done night by night and by considering all the original spectra.

\begin{figure*}
\centering
\includegraphics[width=0.97\textwidth]{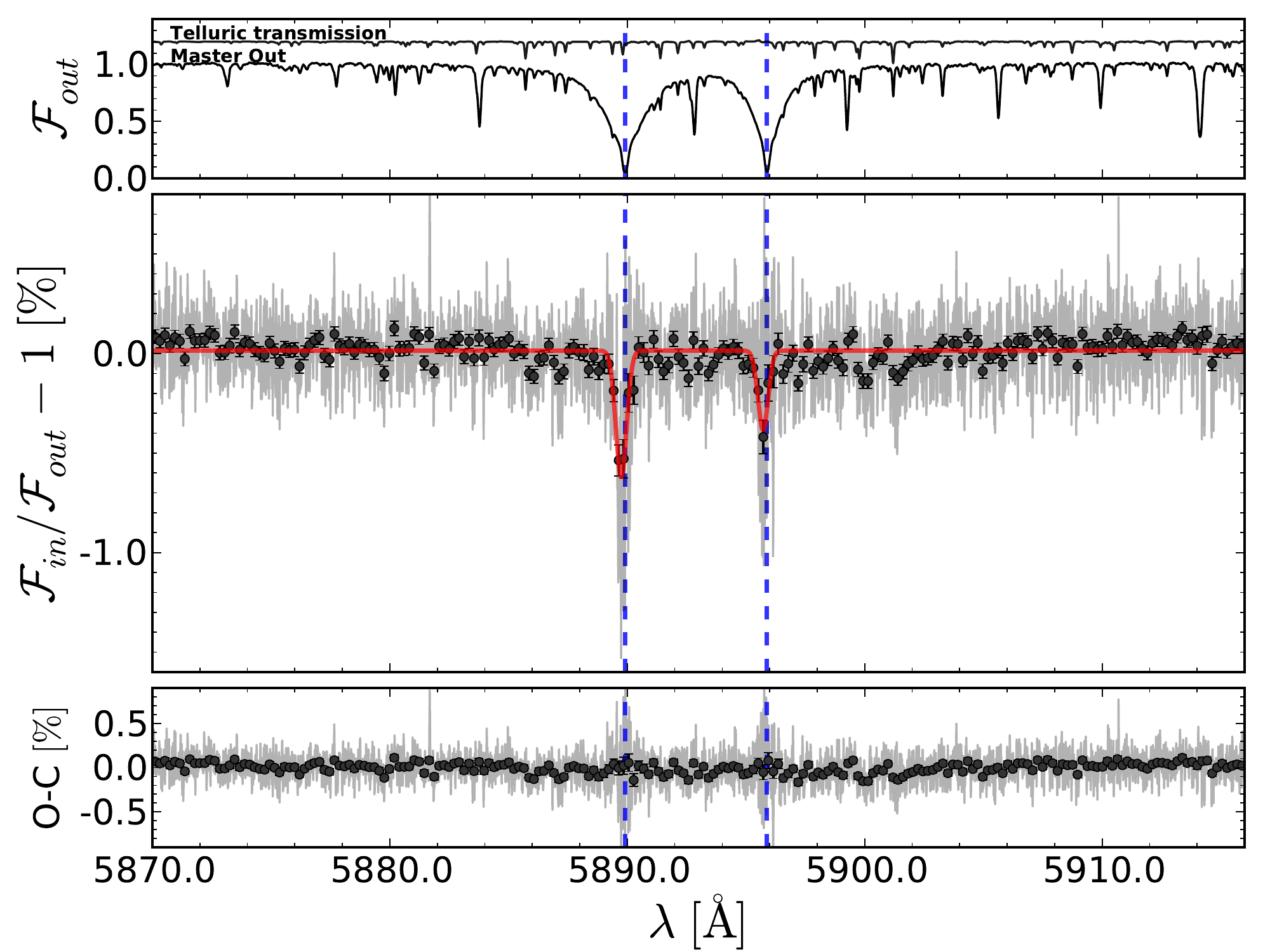}
\caption{HARPS transmission spectrum of HD\,189733b around the sodium  \ion{Na}{i} D doublet. Upper panel: Master out spectrum (stack of the out-of-transit spectra) normalized to unity. A telluric spectrum is also shown with a vertical offset to be able to identify water telluric lines. Middle panel: Overall transmission spectrum of the exoplanet atmosphere taking into account all the observing nights (light grey). The transmission spectrum is shown binned by 20$\times$ with black circles. The planetary radial-velocity correction is applied. We can easily see the two \ion{Na}{i} D lines core absorption from the planetary atmosphere. We show a gaussian fit to each \ion{Na}{i} D lines (red). We measure line contrasts of $0.64\pm0.07\%$ (D2) and $0.40\pm0.07\%$ (D1) and FWHMs of $0.52\pm0.08~\AA$. The transition wavelengths of the doublet, in the planet rest frame, are indicated with blue dashed line. A net blueshift is measured, by our fit, with a value of $0.16\pm0.04\ \AA$. This blueshift may be due to wind in the upper layer of the atmosphere with a velocity of $8\pm2$ km/s (see text). Lower panel: Residuals to the gaussian fit. The noisy part is centered on the two sodium lines wavelength where the stellar flux is the lowest.}
\label{TSpectrum}
\end{figure*}

\section{Results and analysis}\label{Sec_Results}

\subsection{\ion{Na}{i} D detection}\label{Sec_SodiumD}
As we focus on the two transitions of the sodium doublet (D2 at $\lambda$5889.951 $\&$ D1 at $\lambda$5895.924~$\AA$), we restrict the wavelength range for our analysis to the domain from 5870 to 5916~$\AA$. Taking into account the systemic velocity of -2.2765~\kms, these lines are, in the solar system barycentric rest frame, at $\lambda$5889.906 and $\lambda$5895.879~$\AA$, respectively.

To compute absorption depths, we define different central passbands (C) composed by two sub-bands centered on each sodium transition lines. Total band width of 6=2$\times3$~$\AA$, 3=2$\times1.5$~$\AA$, 1.5=2$\times0.75$~$\AA$, 0.75=2$\times0.375$~$\AA$ and 0.375=2$\times0.188$~$\AA$ were chosen (the 6=2$\times3$~$\AA$ is actually 2$\times2.98~\AA$). We add to these bands a larger passband of 12~$\AA$ to perform the comparison of our results to those of \citet{Redfield2008,Jensen2011,Huitson2012}. As the 12~$\AA$ band encompasses the two sodium lines, we adjusted it on the center of the doublet.
For every bands, we choose a unique reference passband corresponding to the blue (B) and red (R) adjacent passbands of the central passband of 12~$\AA$ (i.e. $B=5874.89-5886.89\ \AA$, $R=5898.89-5910.89\ \AA$).

\subsection{Transmission spectrum analysis}
\begin{figure*}
\centering
\includegraphics[width=0.47\textwidth]{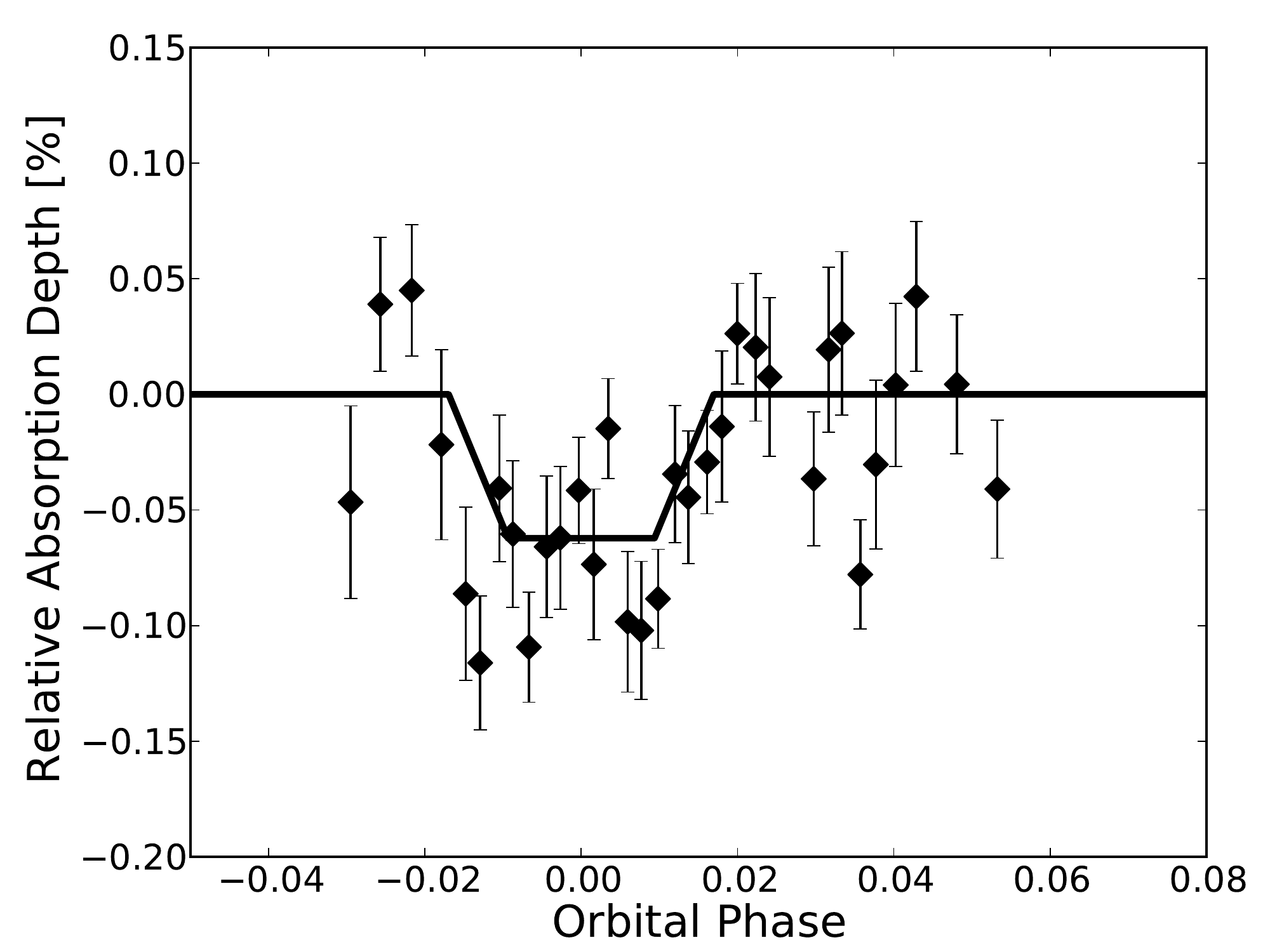}
\includegraphics[width=0.47\textwidth]{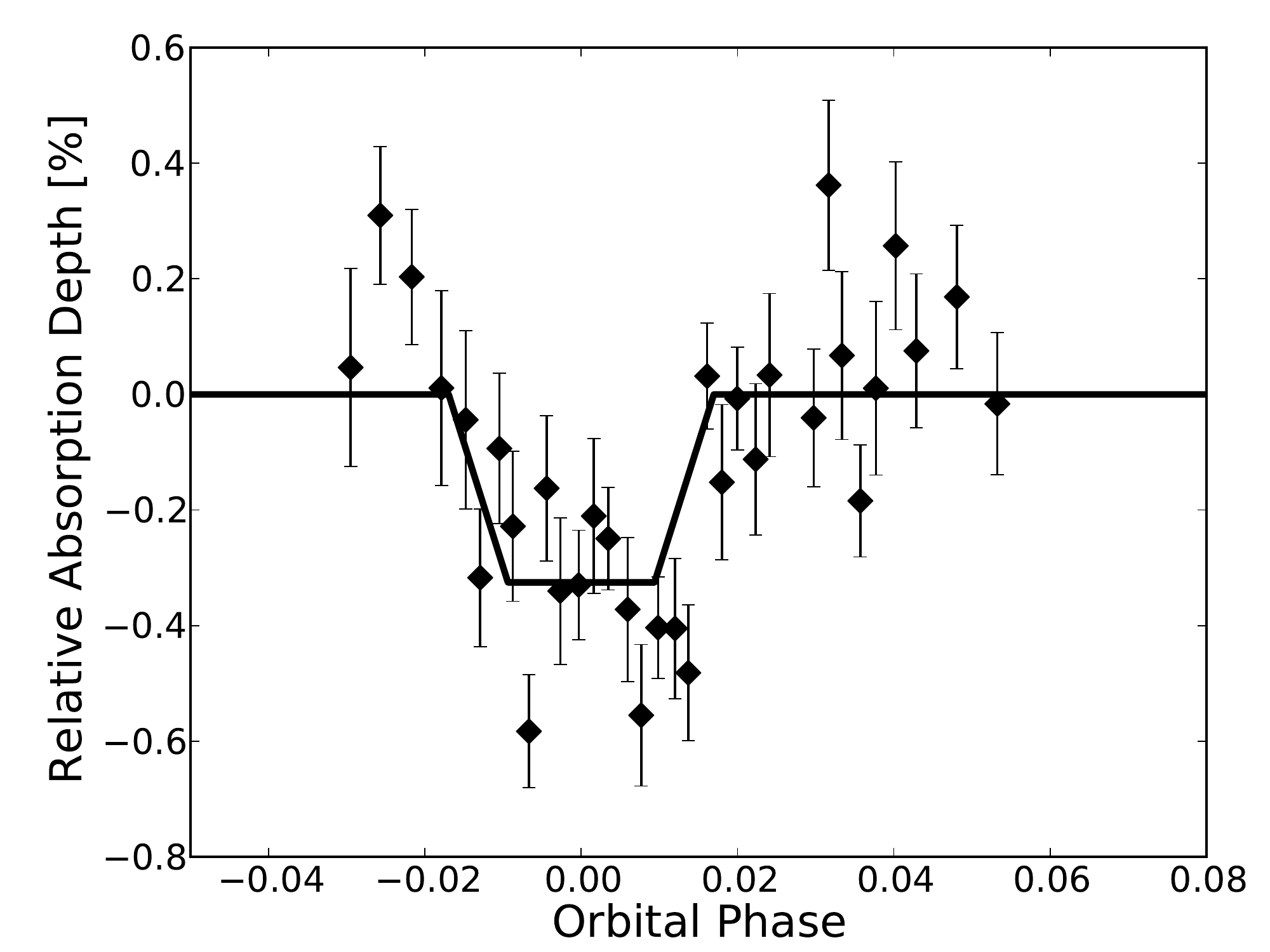}
\caption{HARPS transmission light curve of HD\,189733b. The three nights are averaged together with bins of 3 spectra. Left: The relatives fluxes are integrated in each spectrum over a 12~$\AA$ passband including the whole sodium \ion{Na}{i} D doublet and compared to reference passbands in the same spectrum. The overall absorption depth is detected at 620 ppm with 7.8~$\sigma$. Right: The relatives fluxes are integrated over two passbands of 0.75~$\AA$ around each sodium \ion{Na}{i} D lines. The absorption depth is detected at 3\,250 ppm with 9.8~$\sigma$. Note the different vertical scales.}
\label{TSlightcurve}
\end{figure*}

The total signal-to-noise ratio (SNR) per extracted pixel of the “in” and “out” master spectra ($\mathcal{F}_{in}$ and $\mathcal{F}_{out}$) ranges from 400 to 550 for all nights, in the continuum around 589~nm. A pixel on the HARPS detector represents $\sim$~0.8~\kms\ (or 0.016~$\AA$ at 589~nm). The total SNR for each master is about 850 when co-adding the nights. Thus, a final SNR on the transmission spectra ($\mathfrak{R}^{'}$) of about 1\,200 per extracted pixel is reached. For the transmission spectra, the SNR in the core of the sodium lines decreases to $\sim$~260. It is therefore important to co-add the fluxes over tens to hundreds pixels to reach a sufficient SNR for atmospheric detections. We compute different transmission spectra, notably with and without planetary radial-velocity correction. The standard deviation of the transmission spectra was about 2\,200 ppm for each night in the continuum. When adding the nights together the standard deviation goes down to 1\,450 ppm. As we can see in Fig. \ref{TSpectrum}, the two exoplanetary sodium lines peak out from the continuum. Without planetary radial-velocity correction, the atmospheric absorption is best detected in our 1.5=2$\times0.75~\AA$ passband at a level of $0.309\pm0.034\%$ (9~$\sigma$). Our detection in the 12~$\AA$ passband is measured at a level of $0.060\pm0.008\%$ (7.5~$\sigma$). Here and in the following, we define absorption as a negative relative depth. With planetary radial-velocity correction the signal in these same 1.5=2$\times0.75$ and 12~$\AA$ passbands are $0.320\pm0.031\%$ (10.3~$\sigma$) and $0.056\pm0.007\%$ (8~$\sigma$) respectively (see Table~\ref{table3}). As the radial-velocity correction consists of a wavelength shift smaller than 0.25~$\AA$, it is normal that the integrated signal on these passbands are consistent with each other. Furthermore, our measurements are in agreement with the detection by \citet{Huitson2012} from space ($0.051\pm0.006\%$ for the 12~$\AA$ passbands) and by \citet{Redfield2008} and \citet{Jensen2011} from ground ($0.067\pm0.020$ and $0.053\pm0.017$ for the 12~$\AA$ passband, respectively).

These absorption signals can be interpreted as equivalent relative altitudes. Altitudes can be inferred by considering the transmission at the limb. Indeed, the atmospheric absorption (which depends on wavelength) is due to an optically thick layer (presence of an absorber) at a certain height above the measured planetary radius (in broad band light curve). The height of this layer is generally a few times bigger than the atmospheric scale height of the planet ($H=kT/\mu g$), because of the variation with wavelength of the cross section of the absorber \citep{Lecavelier2008}. The absorption depths measured in the 1.5=2$\times0.75~\AA$ passband correspond to equivalent altitude of $5100\pm500$ km assuming unresolved features. This is about 27 atmospheric scale height of HD\,189733b ($H=190$ km for 1\,140~K, the equilibrium temperature of HD\,189733b, assuming a mean molecular weight of $\mu=2.3$ and an albedo of 0.2).

In order to understand the impact of our radial-velocity correction it is then worth to look at the smallest passband. At first glance, integrations over 0.75=2$\times0.375$~$\AA$ and 0.375=2$\times0.188$~$\AA$ passbands give us smaller signals than the signals we acquire without radial-velocity correction. This is not expected, since the radial-velocity correction should add up all the absorptions signals and thus strengthen the signal. A closer look to the transmission spectrum with a Gaussian fitting to each line shows that the exoplanetary sodium lines are blueshifted by about $0.16\ \AA$ in the planet rest frame. Integrating the lines including this shift (with the smallest passbands blueshifted) enables the full signal to be recovered.

What can produce the blueshift of the exoplanetary sodium lines we measure and is it significant? A global error on the planetary radial-velocity shift is likely to be lower than 3~\kms\ (0.06~$\AA$) considering the HARPS precision on the stellar radial velocity propagated to the planet. Our $0.16\ \AA$ value is by far smaller than the $\sim0.75\ \AA$ measured by \citet{Redfield2008}. Since this specific blueshift would have been easily measured with our data, we do not confirm the value reported by these authors. Nevertheless, our detected $0.16\pm0.04\ \AA$ shift is significant and most likely real, since it is seen even on the transmission spectra without radial velocity correction. A possible physical explanation is that a net blueshift is imprinted by winds within the exoplanet atmosphere \citep[see][for a first detection of winds in exoplanets]{Snellen2010}. Theoretical models \citep[\textit{e.g.}][]{Kempton2014, Showman2013} indeed predict a net blueshift due to both planetary rotation and winds in the terminator of the atmosphere (the part we are probing with transmission spectroscopy). The blueshift we measured corresponds to a wind speed of $8\pm2$ km/s. Models typically predict a 3 km/s blueshift. However, these winds are estimated for pressures down to $\sim10^{-5}-10^{-6}$ bar. As we will see later, our sodium detection allows us to probe pressures of $\sim10^{-7}-10^{-9}$ bar ($0.1-0.001\mu$bar) at altitude of 10\,000~km. We can therefore expect stronger winds than 3 km/s in the higher atmosphere. This calls for additional theoretical works on winds in the upper atmospheres of exoplanets, and new data to confirm the blueshift value.

\begin{figure*}[t!]
\centering{
\includegraphics[width=0.33\textwidth]{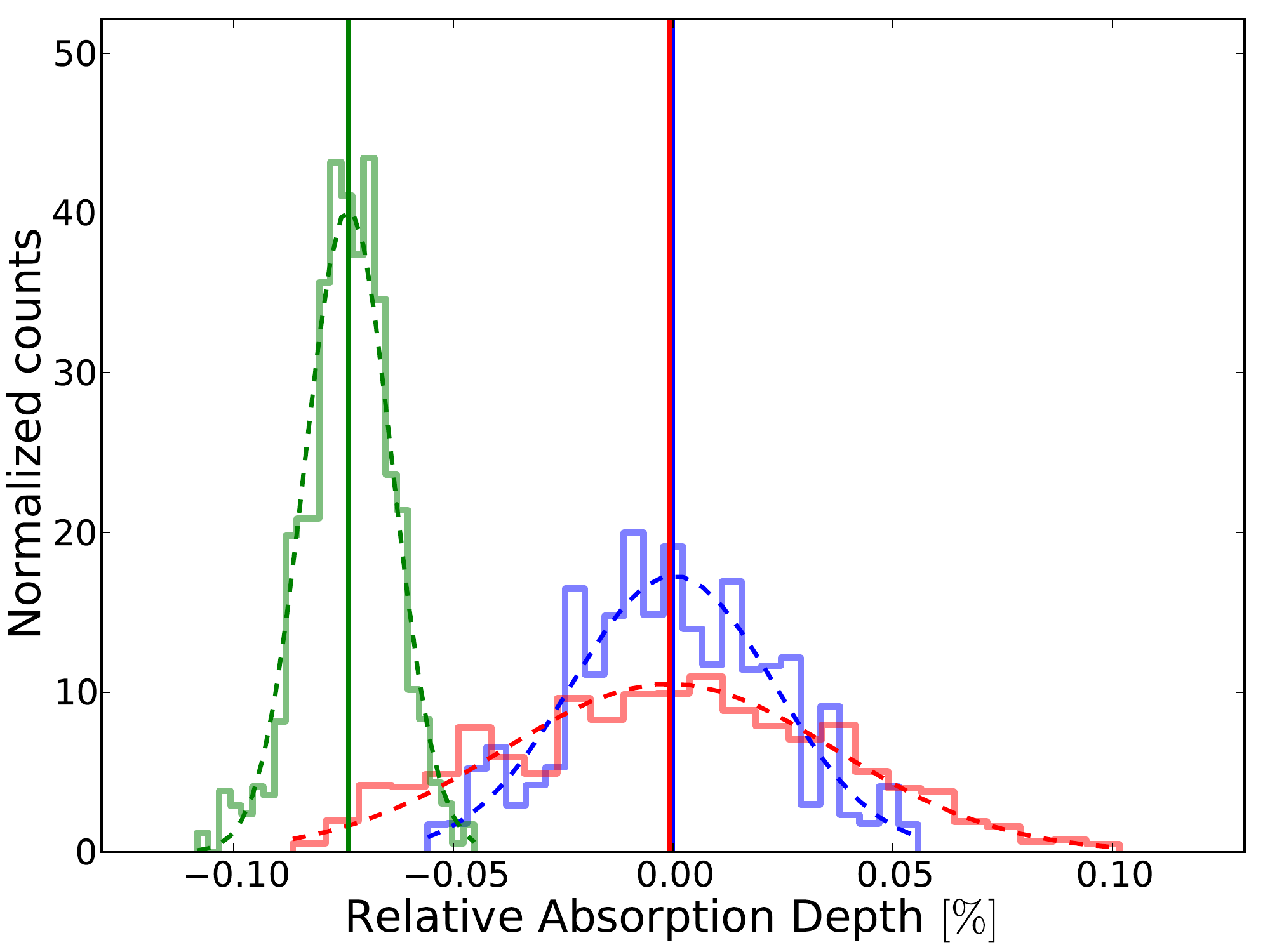}
\includegraphics[width=0.33\textwidth]{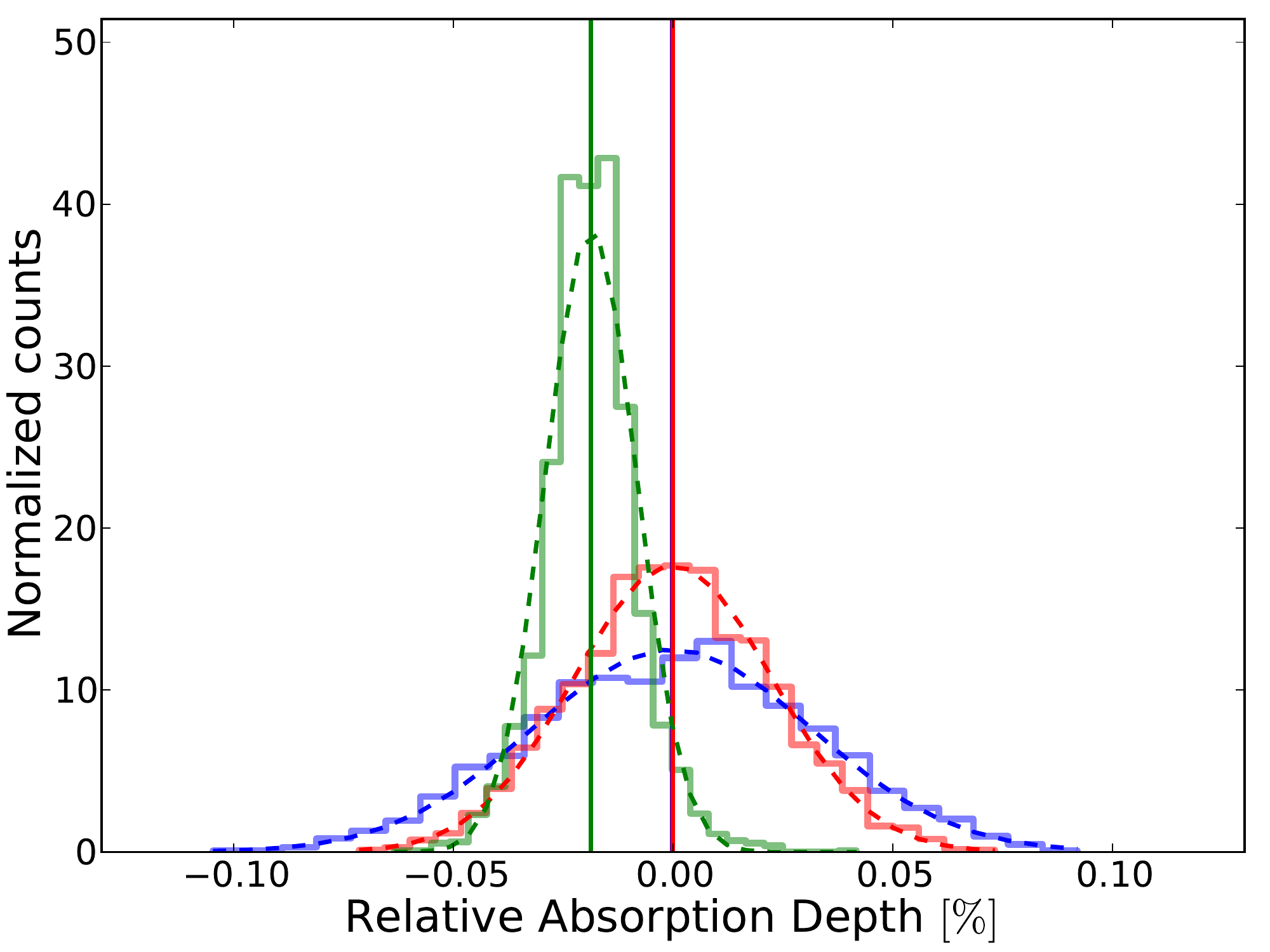}
\includegraphics[width=0.33\textwidth]{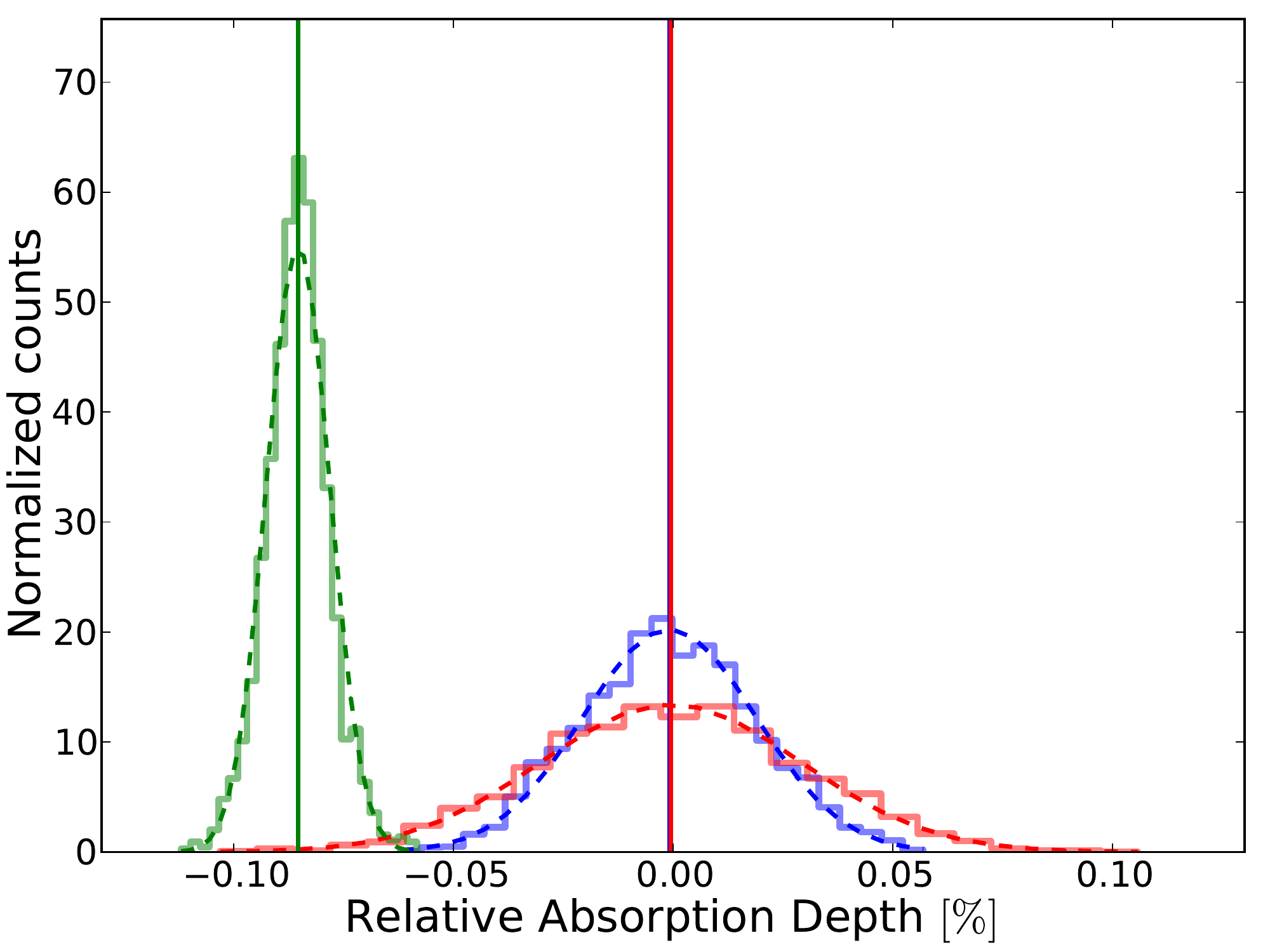}
\includegraphics[width=0.33\textwidth]{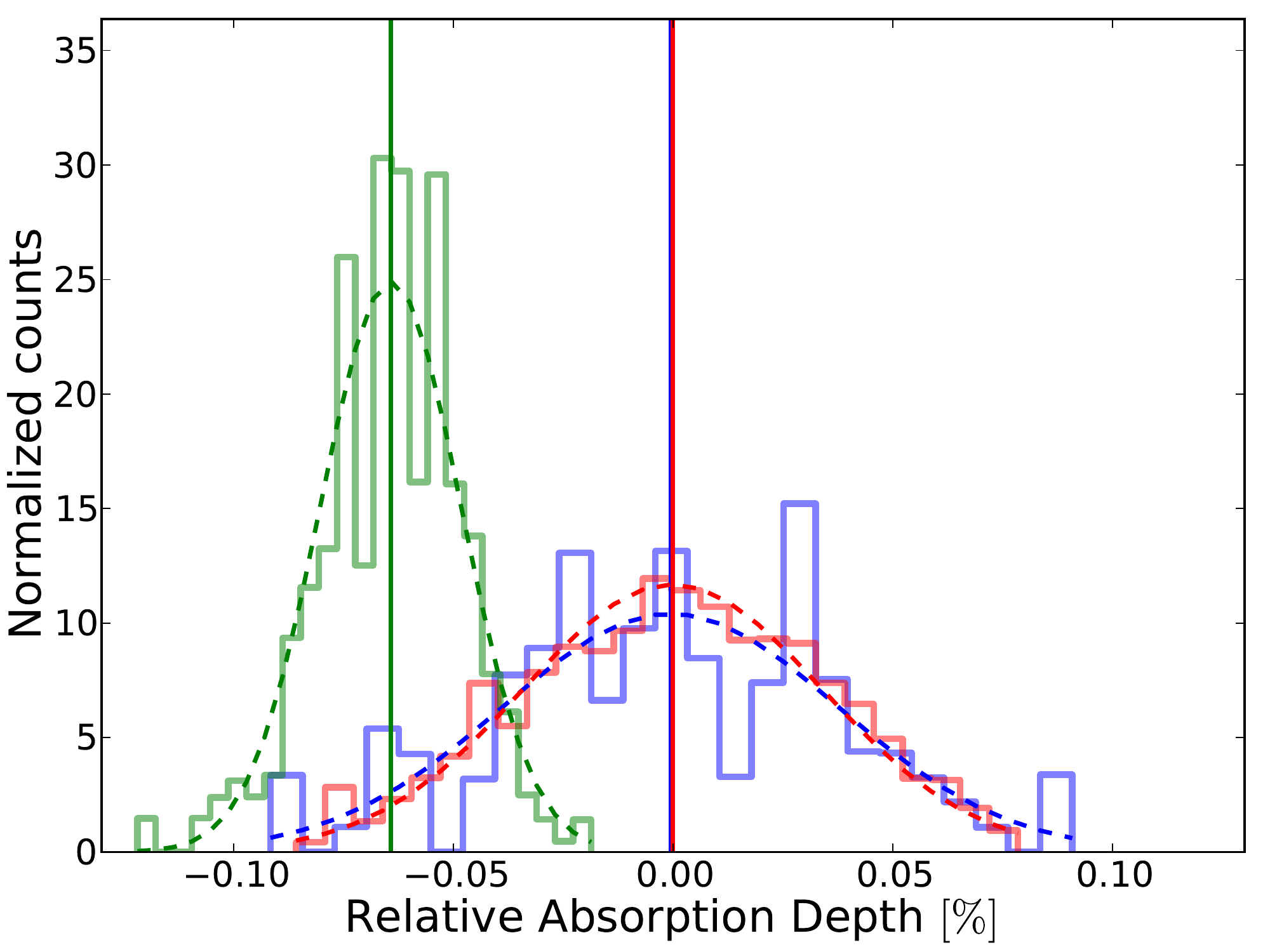}
\includegraphics[width=0.33\textwidth]{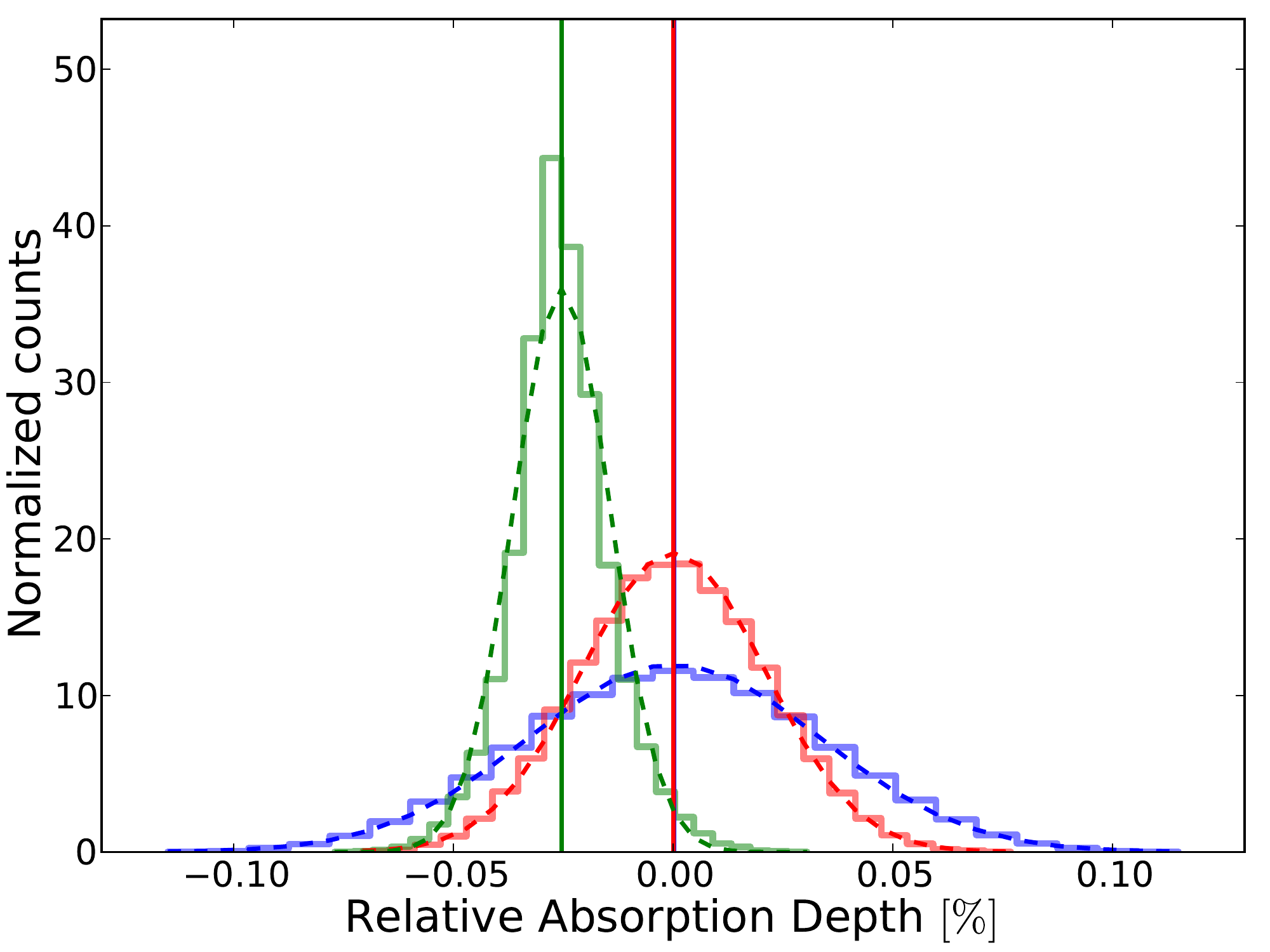}
\includegraphics[width=0.33\textwidth]{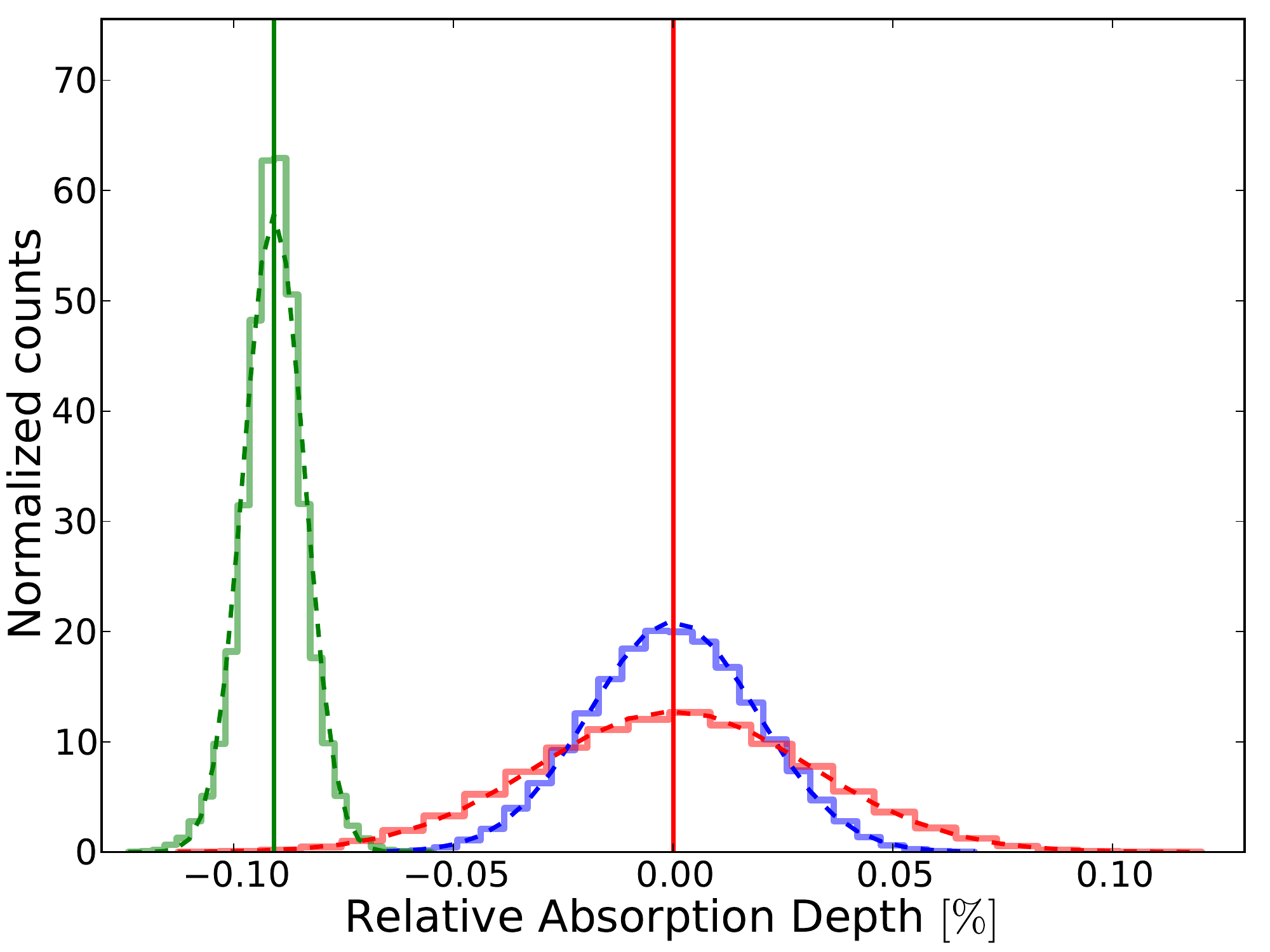}}
\caption{Distributions of the empirical Monte-Carlo analysis for the 12~$\AA$ passband. Top row: results for the transmission spectrum method. Bottom row: results for the light curve method. The columns correspond to the three different nights of observation (nights 1 to 3, from left to right). “in-in”, “out-out” and “in-out” scenarios are shown in blue, red and green. Distributions are shown in continuous lines, Gaussian fit to the distribution in dotted lines and vertical lines are the average values. Results on the measurements and errors coming from these distributions are summarized in Fig.~\ref{TScompar} where we can compare it to the nominal values.}
\label{StatFig}
\end{figure*}

\begin{figure*}[t!]
\centering{
\includegraphics[width=0.33\textwidth]{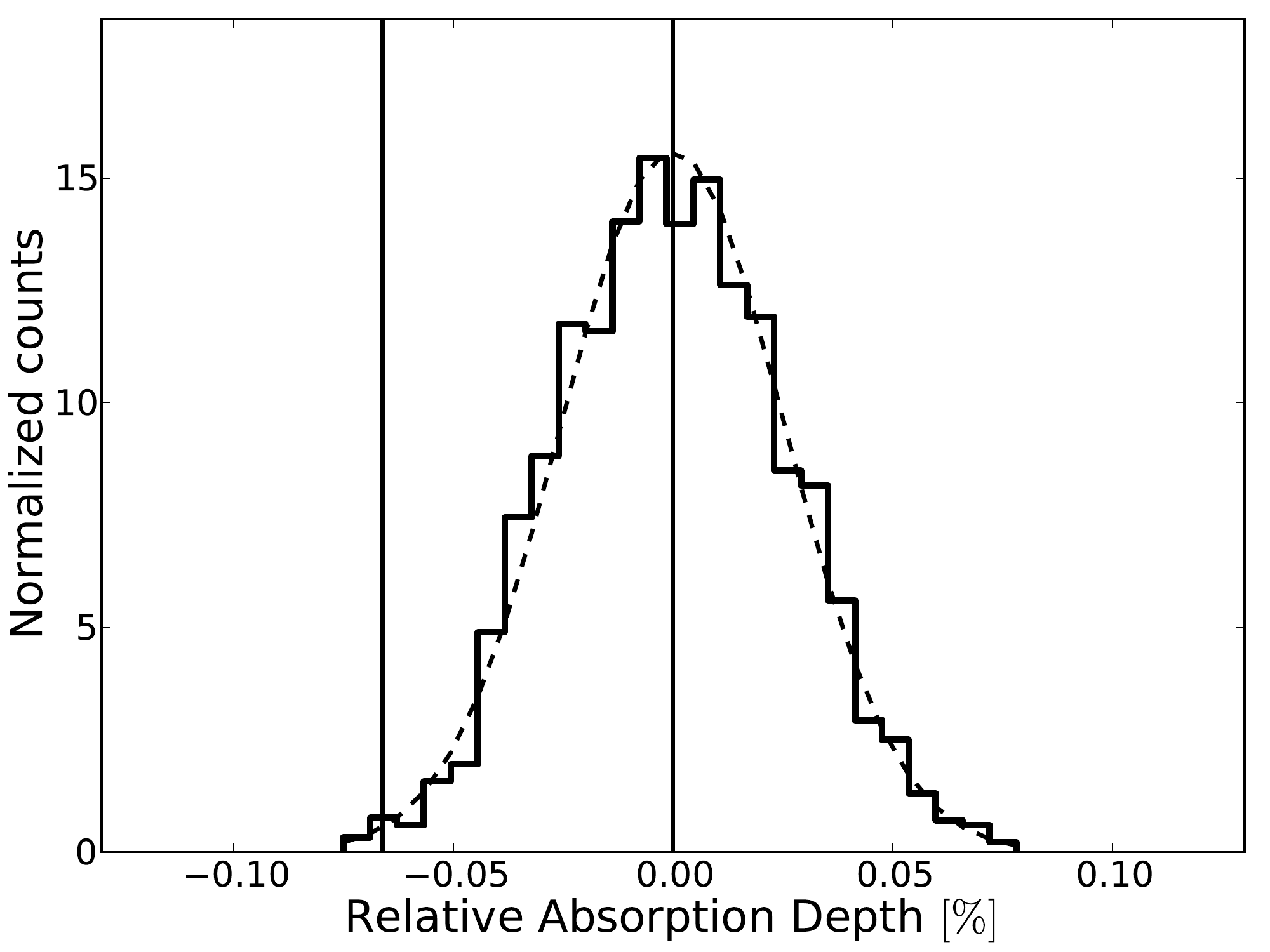}
\includegraphics[width=0.33\textwidth]{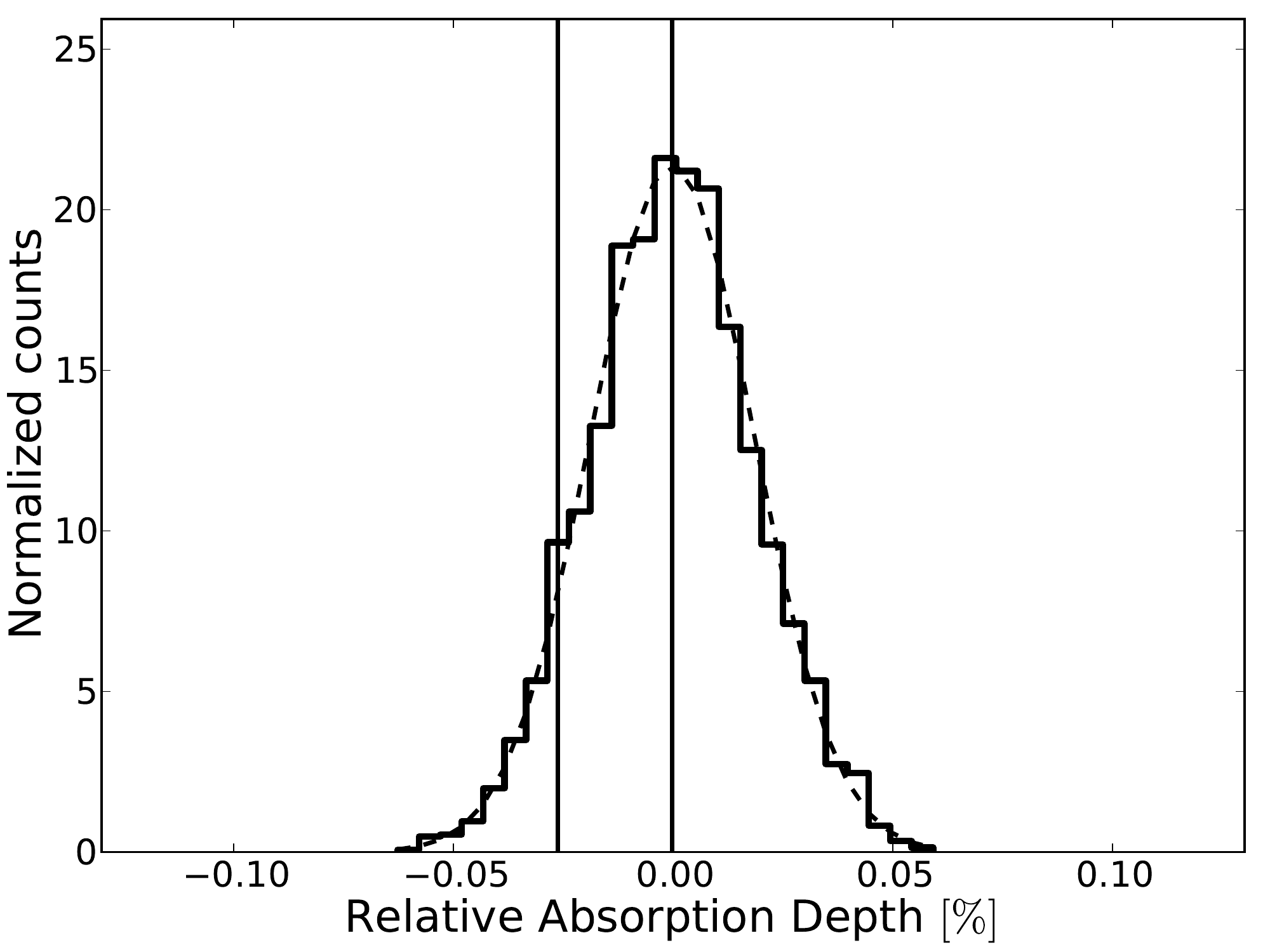}
\includegraphics[width=0.33\textwidth]{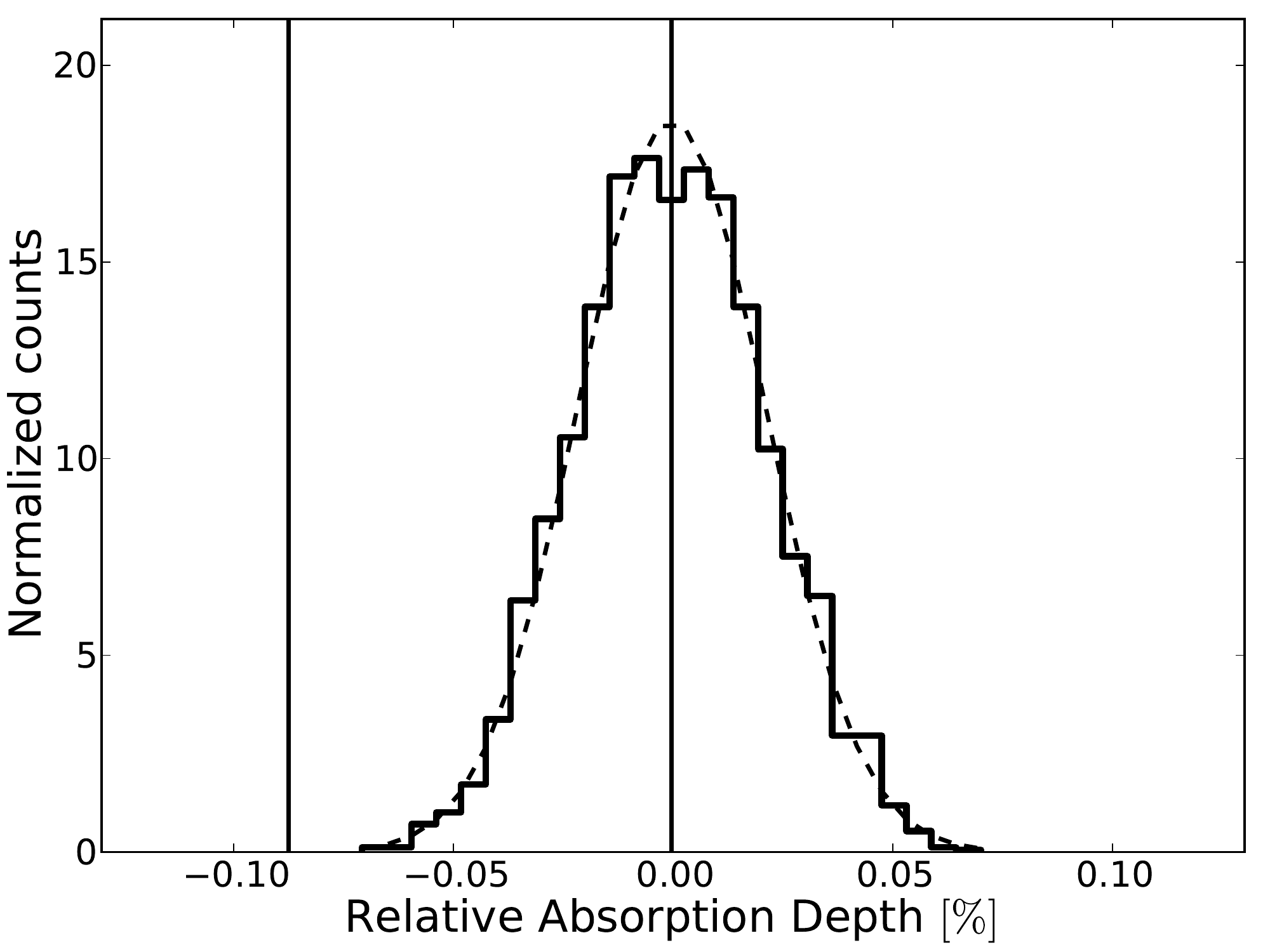}
\includegraphics[width=0.33\textwidth]{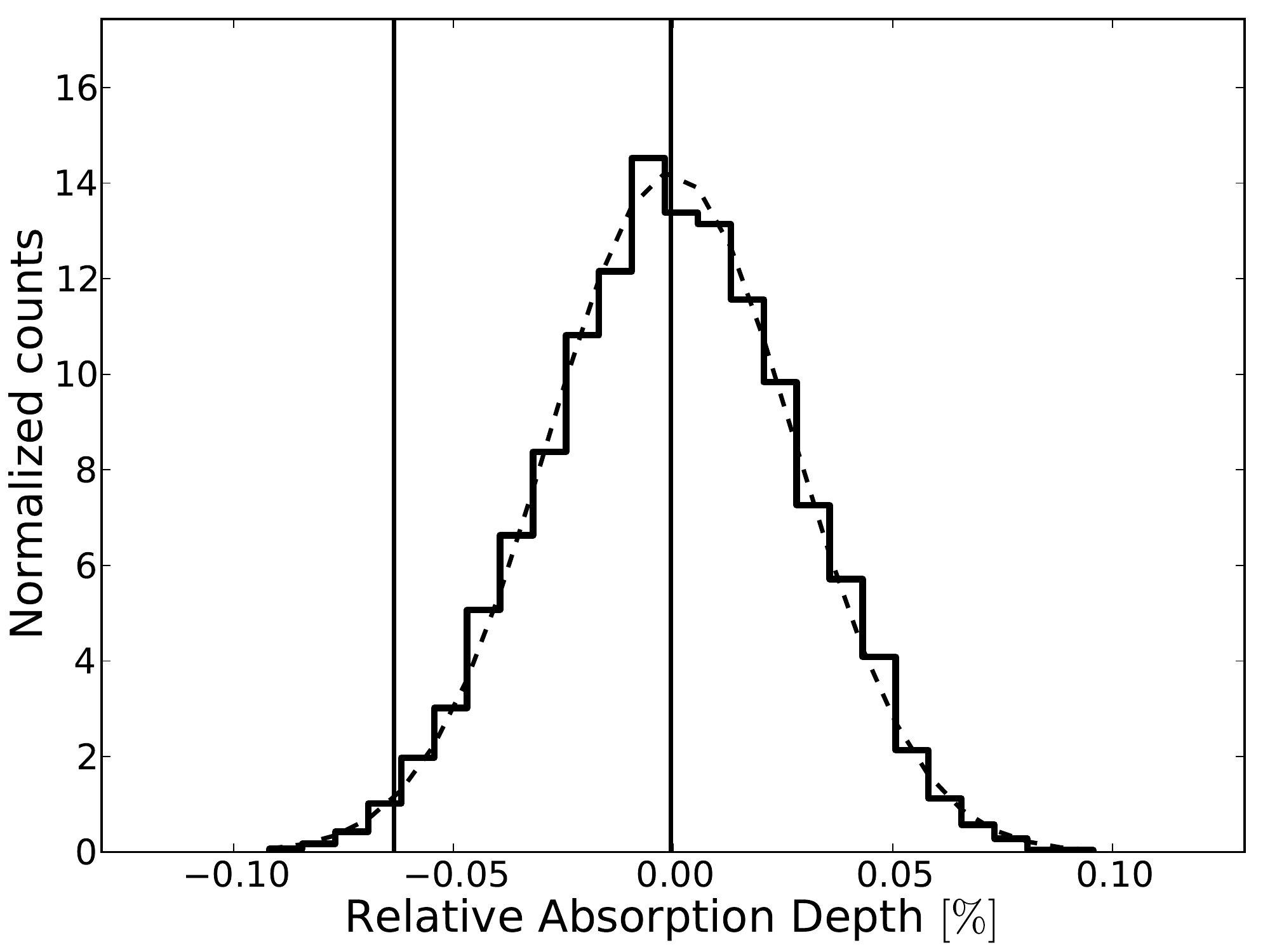}
\includegraphics[width=0.33\textwidth]{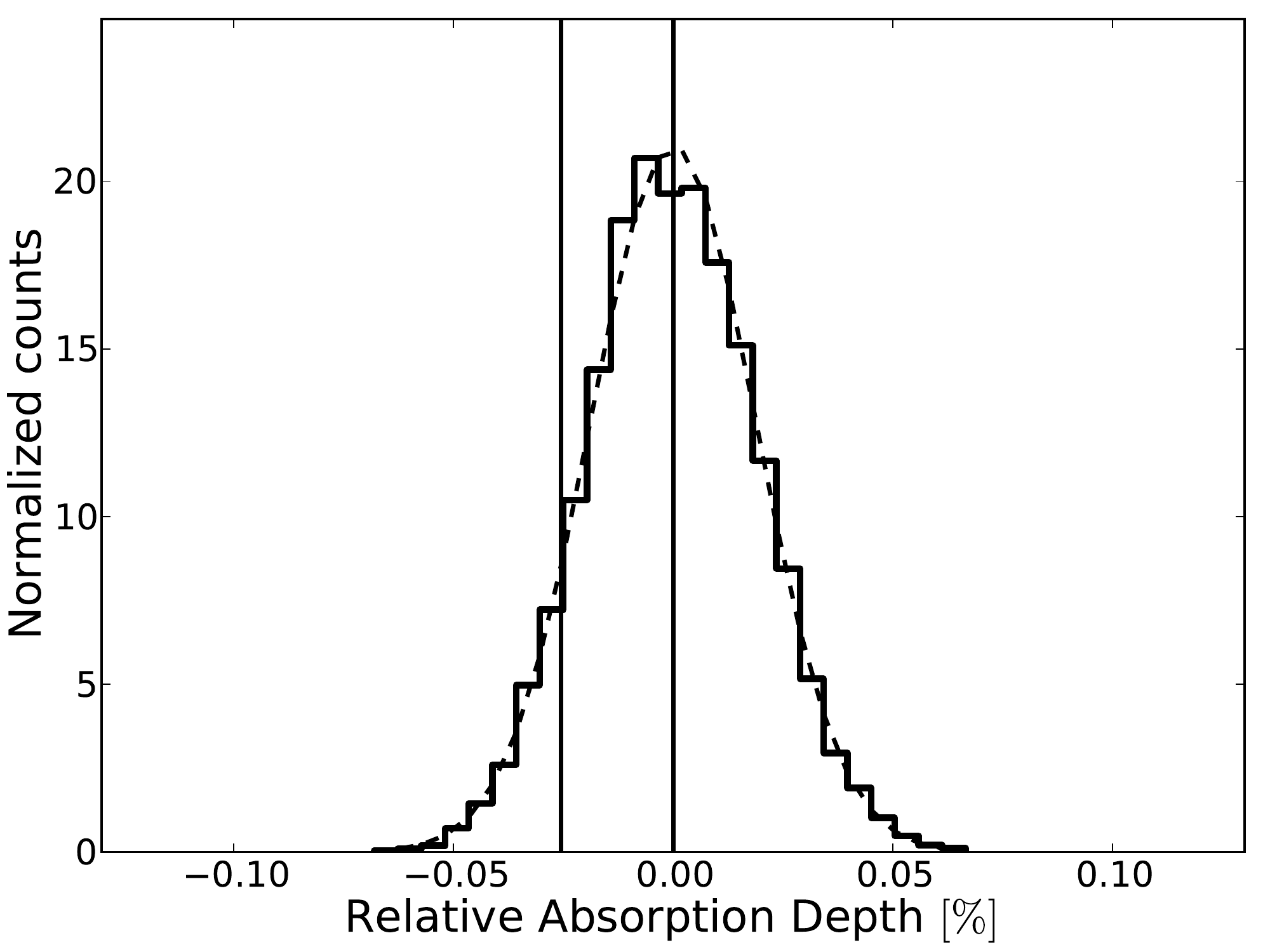}
\includegraphics[width=0.33\textwidth]{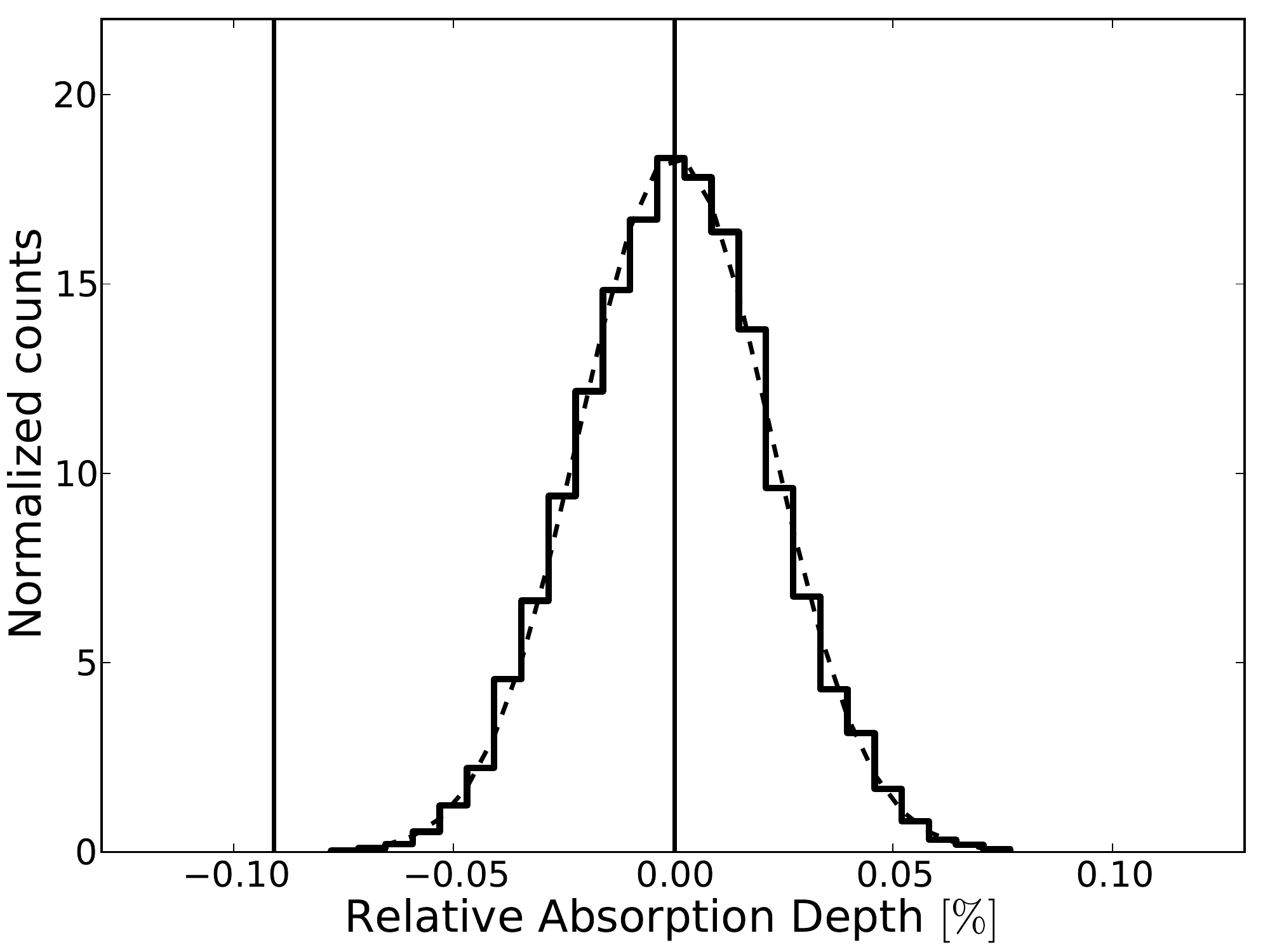}}
\caption{Distributions of the false-alarm probability analysis for the 12~$\AA$ passband. Top row: results for the transmission spectrum method. Bottom row: results for the light curve method. The columns correspond to the three different nights of observation (night 1 to 3, from left to right). Distributions are shown in continuous lines, gaussian fit in dotted lines and nominal values to infer false-alarm probabilities in vertical lines.}
\label{StatFig2}
\end{figure*}

\begin{figure*}[t!]
\centering
\includegraphics[width=0.47\textwidth]{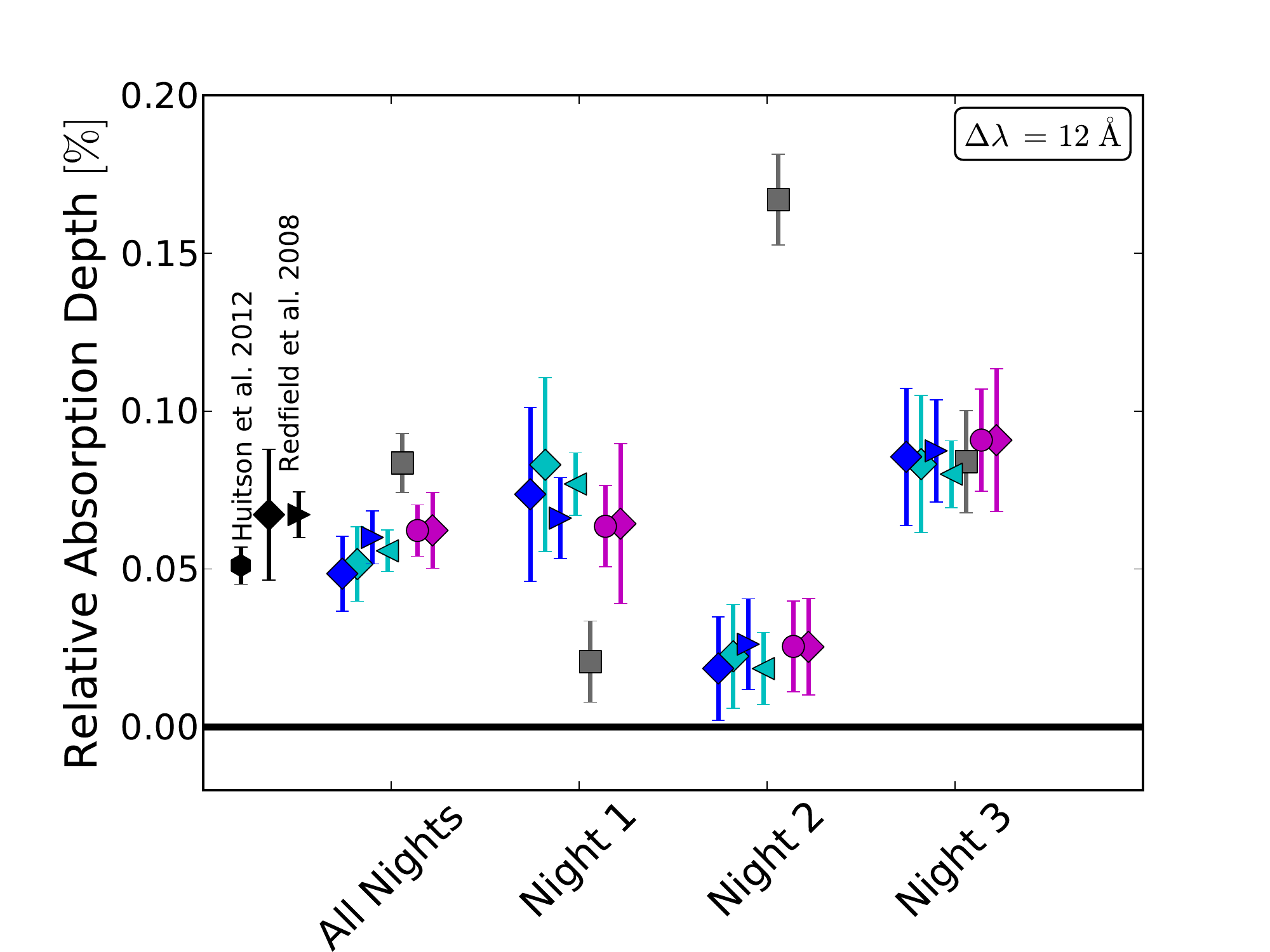}
\includegraphics[width=0.47\textwidth]{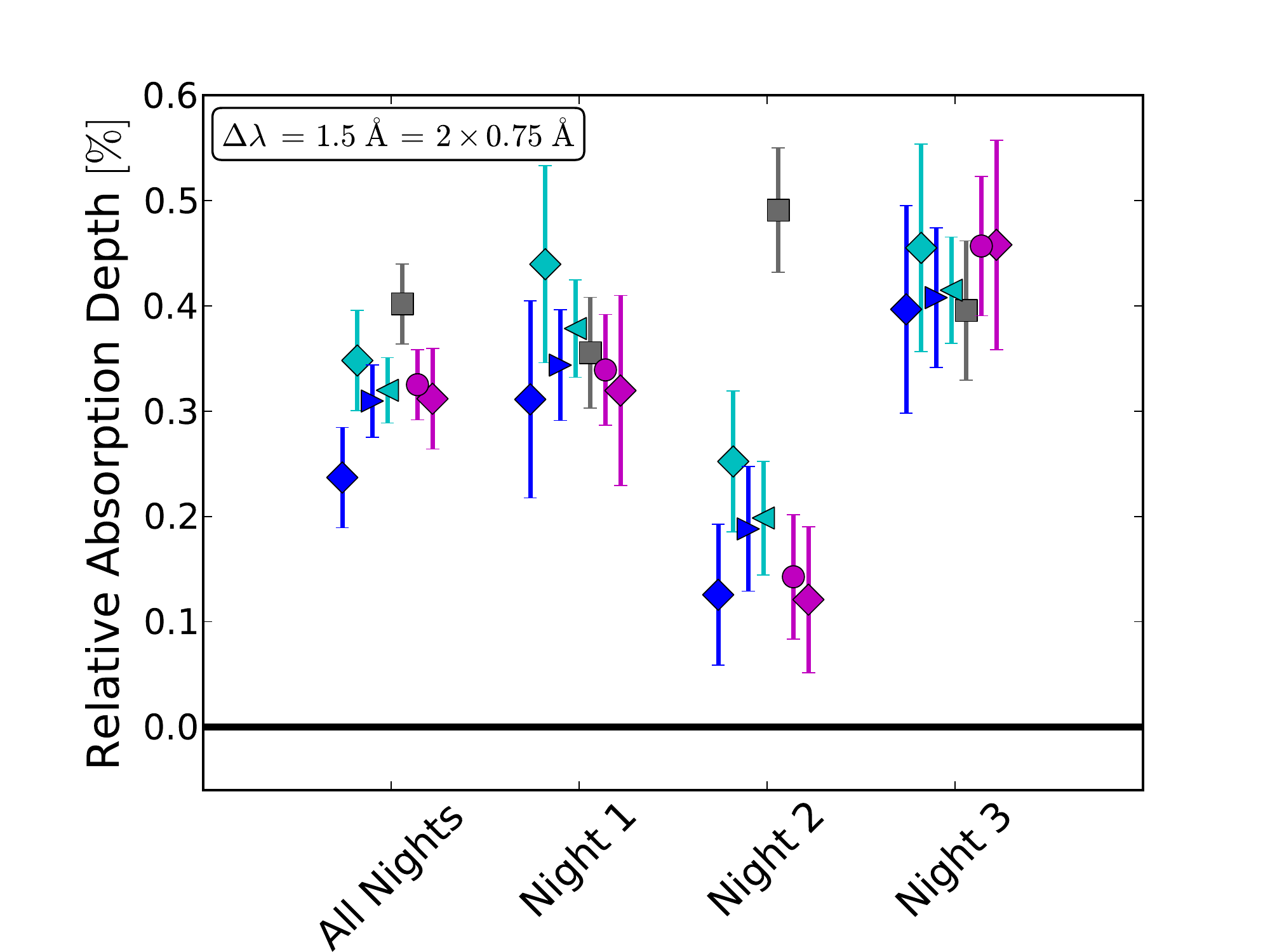}
\caption{Comparison of the different methods. The relative absorption depth due to sodium lines are shown for all the nights and for the different methods. Grey squares are the measurement levels~\textit{without} any correction. Particularly, telluric lines imprint spurious systematics in any directions. Magenta circles and diamonds are measurements made with the light curve method and the empirical Monte-Carlo. Blue left triangle and diamond are for the transmission spectrum without radial-velocity correction and the corresponding empirical Monte-Carlo. Cyan right triangles and diamonds are for the transmission spectrum with planetary radial-velocity correction and the corresponding empirical Monte-Carlo. Total absorption depth when adding all the night are also indicated. Left: Summary for the 12~$\AA$ passband. The measurement of \citet{Huitson2012} (3 transits with HST/STIS) is shown through black hexagon. The measurement of \citet{Redfield2008} (35 exposures in-transit (600~s) observed during 11 transit events) is shown through black left triangle and diamond. The errors correspond to their photon noise and empirical Monte-Carlo analysis, respectively. HARPS precision is comparable with HST/STIS precision showing the importance of ground-based high resolution observation of exoplanetary atmospheres. Right: Summary for the $1.5=2\times0.75\ \AA$ passband.}
\label{TScompar}
\end{figure*}

\begin{table*}[t!]
\caption{Summary of the measured relative absorption depth in [$\%$] on the transmission spectra with radial-velocity correction.}
\begin{center}
\begin{tabular}{lcccccc}
\hline
\rule[0mm]{0mm}{5mm}$\Delta\lambda~[\AA]$ & 0.375=2$\times0.188$&0.75=2$\times0.375$&1.5=2$\times0.75$&3=2$\times1.5$&6=2$\times3$&12\\
\hline
$\#$ pixel &$\sim2\times12$&$\sim2\times25$&$\sim2\times50$&$\sim2\times98$&$\sim2\times194$&$\sim780$\\
\hline
$\mathrm{Night\ 1}$ &0.764$\pm$0.095&0.532$\pm$0.064&0.378$\pm$0.046&0.202$\pm$0.025&0.115$\pm$0.014&0.077$\pm$0.010\rule[0mm]{0mm}{3mm}\\
$\mathrm{Night\ 2}$ &0.495$\pm$0.112&0.431$\pm$0.077&0.198$\pm$0.054&0.028$\pm$0.029&0.015$\pm$0.017&0.018$\pm$0.011\\
$\mathrm{Night\ 3}$ &0.560$\pm$0.104&0.500$\pm$0.071&0.415$\pm$0.050&0.221$\pm$0.027&0.113$\pm$0.015&0.080$\pm$0.010\\
\hline
$\mathrm{All\ Nights}$ &0.571$\pm$0.065&0.472$\pm$0.044&0.320$\pm$0.031&0.141$\pm$0.017&0.075$\pm$0.010&0.056$\pm$0.007\\
\hline
\end{tabular}
\end{center}
\label{table3}
\end{table*}

\begin{table*}[t!]
\caption{Summary of the measured relative absorption depth in [$\%$] on the light curve.}
\begin{center}
\begin{tabular}{lcccccc}
\hline
\rule[0mm]{0mm}{5mm}$\Delta\lambda~[\AA]$ & 0.375=2$\times0.188$&0.75=2$\times0.375$&1.5=2$\times0.75$&3=2$\times1.5$&6=2$\times3$&12\\
\hline
$\#$ pixel &$\sim2\times12$&$\sim2\times25$&$\sim2\times50$&$\sim2\times98$&$\sim2\times194$&$\sim780$\\
\hline
$\mathrm{Night\ 1}$ &0.606$\pm$0.179&0.468$\pm$0.095&0.339$\pm$0.053&0.207$\pm$0.031&0.127$\pm$0.018&0.064$\pm$0.013\rule[0mm]{0mm}{3mm}\\
$\mathrm{Night\ 2}$ &0.244$\pm$0.199&0.165$\pm$0.106&0.143$\pm$0.059&0.040$\pm$0.034&0.027$\pm$0.021&0.026$\pm$0.014\\
$\mathrm{Night\ 3}$ &0.521$\pm$0.224&0.661$\pm$0.119&0.457$\pm$0.066&0.274$\pm$0.039&0.133$\pm$0.023&0.091$\pm$0.016\\
\hline
$\mathrm{All\ Nights}$ &0.478$\pm$0.113&0.451$\pm$0.060&0.325$\pm$0.033&0.184$\pm$0.019&0.102$\pm$0.012&0.062$\pm$0.008\\
\hline
\end{tabular}
\end{center}
\label{table4}
\end{table*}

\subsection{Transmission light curve analysis}
We compute the transmission light curve of HD\,189733b for every observation nights following Sect.~\ref{light_curve_method} and for every passbands described in Sect.~\ref{Sec_SodiumD}. For our analysis, the transit ephemeris are fixed. They are given by the parameters described in Table~\ref{table2}. For the 12~$\AA$ passband, we estimate errors on each individual spectrum of 250 ppm to 800 ppm. The propagation of these errors to the time series  provides us uncertainties of about 100--150 ppm on the values of the baseline and on the absorption depth. This allows us to determine absorption depths due to sodium in the exoplanet atmosphere during each night at a level between 1.9 and 5.7~$\sigma$ (see Table~\ref{table4}). Note that when we subtract the best-fit absorption depth, the standard deviation of the residuals is about 550 ppm for each night. When co-adding data of the three nights together, we obtain an absorption signal of $0.062\pm0.008\%$ (7.8~$\sigma$) for the 12~$\AA$ passband (Fig. \ref{TSlightcurve}). Our best measurement is achieved for the passband of $1.5=2\times0.75\ \AA$ and delivers a value of $0.325\pm0.033\%$ (9.8~$\sigma$). This is in perfect agreement with our transmission spectrum method and shows both the complementarity and the compatibility of the two methods. Compatibility because the measured absorption level are within a 1~$\sigma$ error bars \citep[see][]{Astudillo-Defru2013} and complementarity because the second method carries line information and demonstrates that the additional absorption is indeed observed during the transit (see also Sect.~\ref{Stat1}). As expected, a small decrease on the measured absorption signal is observed for the smallest passband (0.375=2$\times0.188$~$\AA$), which width is commensurable with the wavelength shift imprinted by the radial velocity of the planet. This effect thus results in a flux loss, hence to a weaker signal.

\subsection{Systematic effects}\label{Stat1}
Possible systematic errors can be present in the data. To estimate such effects, we perform an empirical Monte-Carlo (EMC) analysis for each of our observed transits following Sect.~\ref{Stat_methods}. This analysis makes us confident that our measured signals is indeed due to the exoplanetary transit and not spurious (\textit{e.g.} due to other astrophysical effects as stellar rotation, observational conditions, or instrumental effects). The EMC is performed for both the transmission spectrum and the light curve methods using 3\,000 iterations for each scenario described in Sect.~\ref{Stat_methods} (“in-in”,“out-out”,“in-out”) and for each night. The standard deviation of the distributions does not increase significantly when increasing the number of iterations.

The results are shown in Fig.~\ref{StatFig} and the errors are summarized with all the different methods in Fig.~\ref{TScompar}. We can see that all the “in-in” and “out-out” distribution are centered on zero showing that the measured absorption signal comes indeed from the transit and has no other sources of explanation. The “in-out” distributions are well centered at the value determined by the transmission spectrum and light curve methods. The errors given by the EMC are larger than the nominal errors (yielding from the propagation of the photon noise), but only by a factor of $\lesssim 2$. This shows that while systematic errors are present in our data they do not undermine our detection. The empirical Monte-Carlo simulation described above strengthens the transit origin of the detected signal.

We also computed false alarm probabilities for each night for both the transmission spectrum and the light curve methods. We performed 10\,000 iterations and found false-alarm probabilities to our sodium detection of $\sim 0.5\%$, $\sim 9\%$ and $\lesssim0.01\%$ for night 1, 2, and 3 respectively (see Fig.~\ref{StatFig2}). Adding all the nights yields a false-alarm probabilities $\lesssim0.01\%$.

\section{Discussion}\label{Sec_Discuss}

\subsection{Summary of the \ion{Na}{i} D detection}
A summary of our results is shown in Fig.~\ref{TScompar}. All the different methods to measure the transit depth $\delta(\Delta\lambda)$ in a given wavelength bin $\Delta\lambda$ are shown for each night individually and also for the combined data set, for the $1.5=2\times0.75$~ $\AA$ and 12~$\AA$ passbands.

A number of aspects must be pointed out: First, the correction of the telluric lines is mandatory. In fact, water lines around the sodium \ion{Na}{i} D doublet imprint spurious systematics in any possible direction. It depends if the transit is observed at low or high airmass compared to the out-of-transit data and at which value of the barycentric Earth radial velocity (BERV). Indeed, due the BERV, the telluric lines will not always fall at the same place in the stellar spectra. For a fixed choice of passbands, a given telluric line may not be always included. Then, the ratio with the reference passbands, which contain other telluric lines, can be affected in any direction. For example, during night 3, the signal measured without telluric correction is, by chance, the same signal measured after telluric correction, as the positions of telluric lines in the transmission spectrum do not impact the sodium signal extraction. It should be noted that the correction for night 2 is less robust than the one of the two other nights, because of the lack of a baseline before the transit. This implies that we probably over-corrected the data taken during this night, hence the measured value is under estimated. Secondly, the absorption depths measured with transmission-spectra and light-curve methods are consistent with each other within the same night. Thirdly, systematic effects are different from night to night. Nights 2 and 3 were probably more “stable” than night 1. Systematics in night 1 are probably also bigger due to the low-cadence of the observations. Even if similar signal-to-noise ratios are reached in the three different nights, it seems that it is better to have a high-cadence observational strategy (provided all the spectra have a SNR above a certain threshold). We also note that the measured values are different for different nights. At this level of precision, and assuming we are efficiently correcting for systematics, we cannot exclude intrinsic variability related to the stellar or planetary properties. Our results favor a variation from night to night within 1--2 $\sigma$. We did not take into account the potential effect of a differential limb darkening or of the Rossiter-McLaughlin effect. However, the impact of differential limb-darkening should be an order of magnitude lower than the achieved precision \citep{Charbonneau2002,Redfield2008}. On the other hand, the Rossiter-McLaughlin effect can mimic an exoplanetary atmospheric absorption and imprint features at similar level in a transmission spectrum. Nonetheless it should average out during a whole transit if the exposures are equally distributed over the transit \citep{Dravins2014}. This is the case for our three analyzed transit events. Finally, we would like to point out that the precision obtained from the coaddition of all the nights is comparable with HST/STIS precision (with the same number of observed transits) or with 10-m sizes telescopes. This highlights the potential of ground-based high resolution observations of exoplanetary atmospheres from 4-m telescopes.

\begin{figure}[t!]
\centering{
\includegraphics[width=0.47\textwidth]{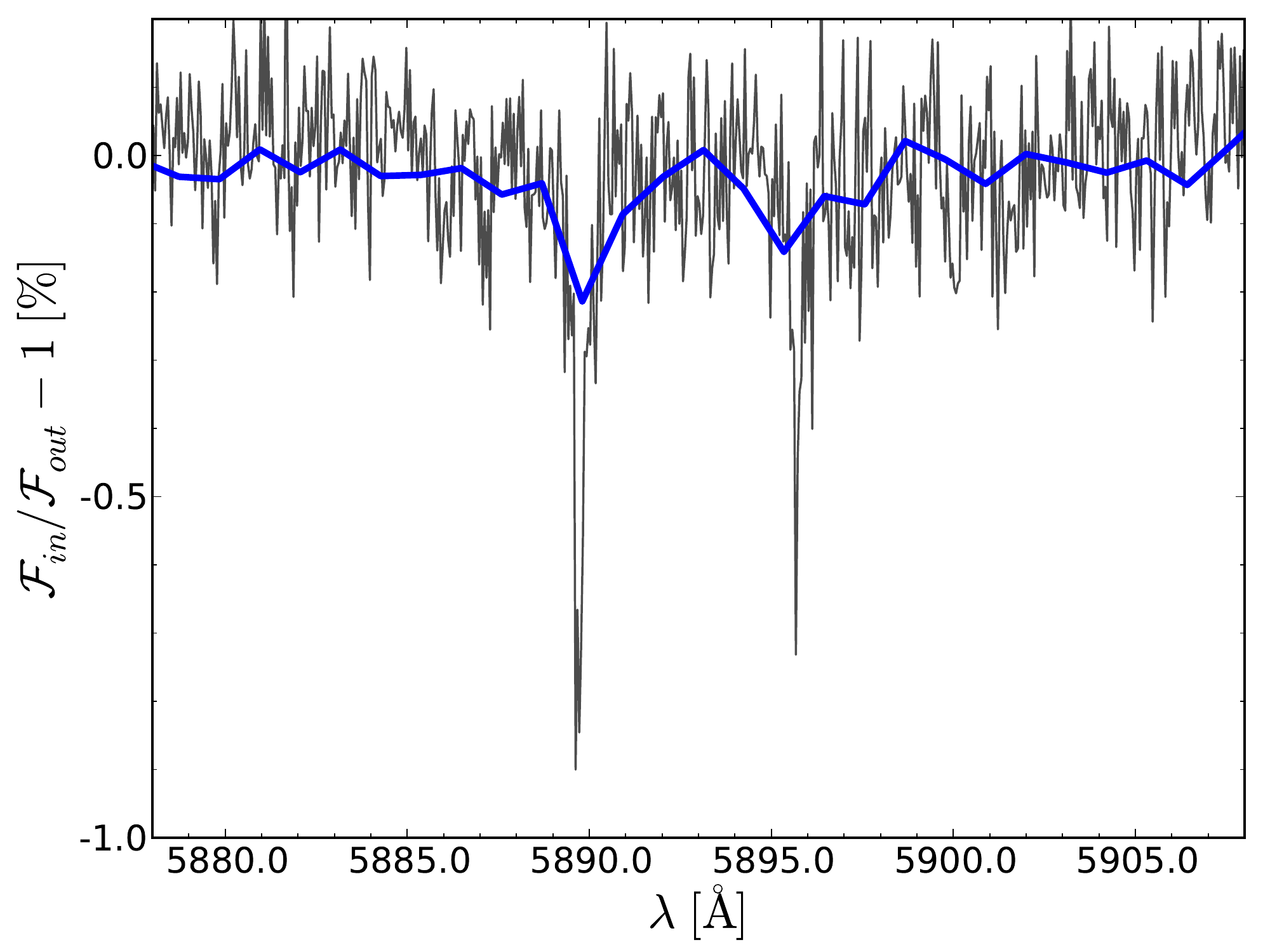}}
\caption{Comparison of the transmission spectra around the \ion{Na}{i} doublet obtained by the HST/STIS instrument \citep[][in blue]{Huitson2012} and by the HARPS spectrograph (binned by 5$\times$, in black).}
\label{ResLine}
\end{figure}

\begin{figure*}[t!]
\centering{
\includegraphics[width=0.97\textwidth]{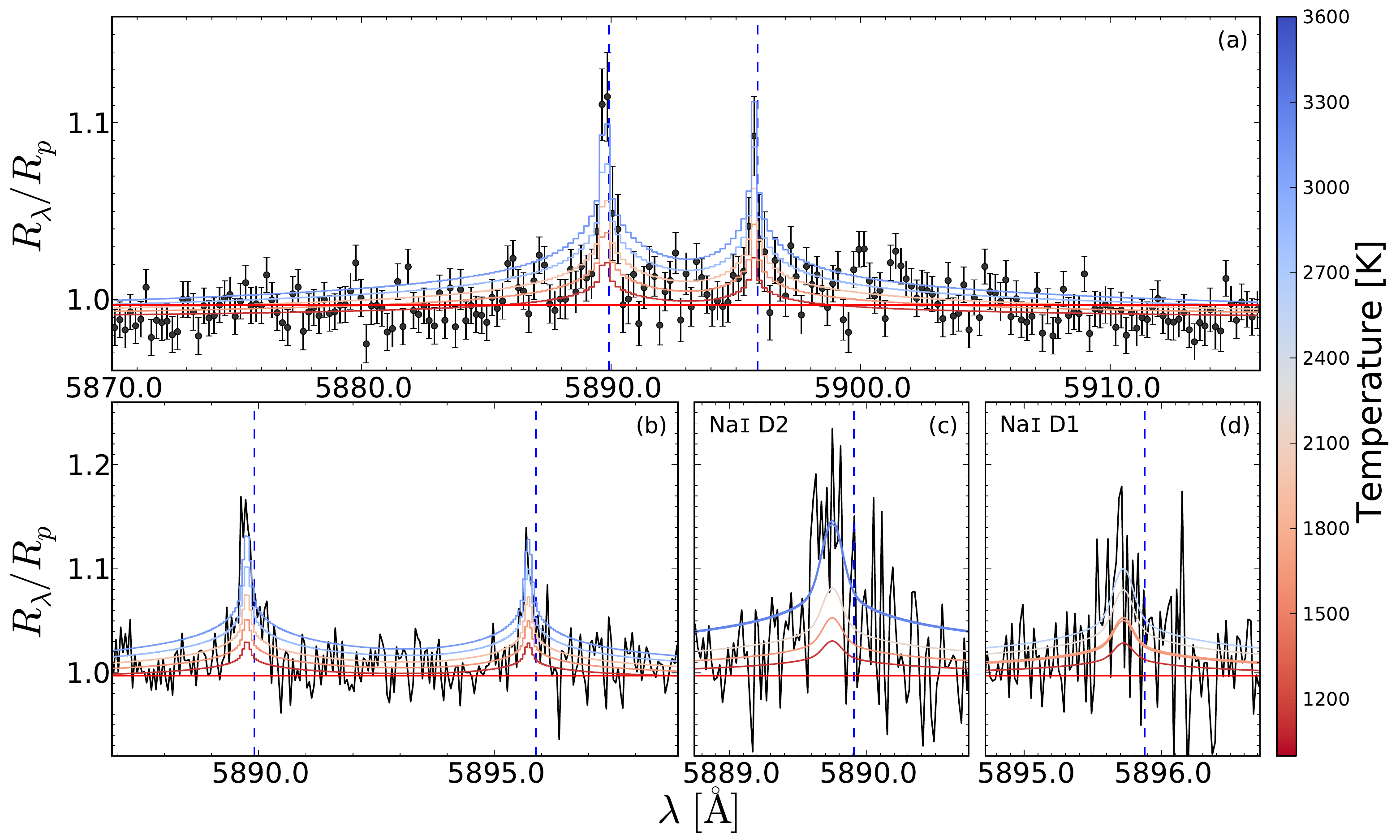}}
\caption{Fit of a range of $\eta$ models to the HARPS transmission spectra of HD\,189733b. The data and the models are binned by 20$\times$ (panel a), 5$\times$ (panel b) and 2$\times$ (panels c and d). Models correspond to those described in Table~\ref{table5}. Some additional models (2100~K, 2600~K, 3100~K, the hottest in panels a and b) and the continuum level (in red), which correspond to the altitudes of a haze layer deck, are also shown for comparison. Our analysis shows that the atmosphere of HD\,189733b is non-isothermal.}
\label{Fit}
\end{figure*}

\begin{table*}[t!]
\caption{Choice of the wavelength ranges for the atmospheric model fits, number of \ion{Na}{i} lines FWHM encompassed in the wavelength ranges, corresponding altitudes, and fitted temperatures. We centered the wavelength ranges at the center of each sodium line in the planet rest frame taking into account the observed blueshift.}
\begin{center}
\begin{tabular}{lcccc}
\hline
\rule[0mm]{0mm}{5mm}$\mathrm{Model\ Atmospheres}$ & Wavelength Ranges [$\AA$] & $\#$ \ion{Na}{i} D's FWHM & Corresponding Altitude [km] & Fitted Temperature [K]  \\
\hline
$\mathrm{Model\ 0}$ & 5870.00--5882.22 & --& 0 & 1\,140\\
 & 5903.24--5916.00 &  &  & \\
\hline
$\mathrm{Model\ 1}$ & 5882.22--5889.22 & 2--14 & 1\,500$\pm$1\,500 & 1\,630$\pm$70\\
$\mathrm{(}$“$\mathrm{Wing'shoulders}$”$\mathrm{)}$ & 5890.26--5895.20 &  & & \\
 & 5896.24--5903.24 & & & \\
$\mathrm{Model\ 2a\ (D1\ core})$ & 5895.20--5895.46  & 1--2 & 2\,700$\pm$800 & 1\,700$\pm$320\\
 & 5895.98--5896.24 & & & \\
$\mathrm{Model\ 2b\ (D2\ core})$ & 5889.22--5889.48 & 1--2 & 3\,800$\pm$900 & 2\,170$\pm$320\\
 & 5890.00--5900.26  & & & \\
$\mathrm{Model\ 3a\ (D1\ core})$ & 5895.46--5895.67 & 0.2--1 & 5\,100$\pm$3\,100 & 2\,220$\pm$340\\
 & 5895.77--5895.98 & & & \\
$\mathrm{Model\ 3b\ (D2\ core})$ & 5889.48--5889.69 & 0.2--1 & 7\,900$\pm$5\,500 & 3\,220$\pm$270\\
 & 5889.79--5890.00 & & & \\
$\mathrm{Model\ 4a\ (D1\ core})$ & 5895.67--5895.77 & $\leqslant0.2$ & 9\,800$\pm$2\,800 & 2\,600$\pm$600\\
$\mathrm{Model\ 4b\ (D2\ core})$ & 5889.69--5889.79 & $\leqslant0.2$ & 12\,700$\pm$2\,600 & 3\,270$\pm$330\\
\hline
\end{tabular}
\end{center}
\label{table5}
\end{table*}

As mentioned before, we obtain the same precision on our detection ($0.056\pm0.007\%$) as on the HST/STIS measurements of \citet{Huitson2012} ($0.051\pm0.006\%$) on a relatively wide passband (here the “12~$\AA$”). We can take advantage of the HARPS high spectral resolution ($\mathcal{R}\sim115\,000$), compared to HST/STIS ($\mathcal{R}\sim5\,500$) to characterize the narrow line cores of the \ion{Na}{i} doublet and, as a consequence, exploring higher regions in the atmosphere of HD\,189733b.

The presence of haze in the HD 189733b atmosphere has been previously deduced from low-resolution spectra covering much broader spectral regions than the one considered here \citep[\textit{e.g.} in][]{Lecavelier2008,Pont2013}. Rayleigh scattering by hazes dominates the transmission spectrum in the optical, only allowing the narrow core of the sodium lines to be observed at high spectral resolution. This is indeed what we observe in our data.

As described in Sect.~\ref{Sec_SodiumD}, we compute different absorption depths $\delta(\Delta\lambda)$ for different passbands (summary in Table~\ref{table3} and~\ref{table4}). The absorption depth increases for narrow cores passbands. This shows that most of the signal we measure originates in the narrow line cores ($\leq1\AA$), even if a small portion of the absorption comes from a broader component. Actually, a Gaussian fitted to each lines yields a full width at half maximum (FWHM) of $0.52\pm0.08~\AA$ (see Fig.~\ref{TSpectrum}). A comparison to the resolution element of 0.05~$\AA$ (2.7~\kms) shows that the lines are resolved by a factor of $\sim10$ (see Fig.\ref{ResLine}). The same fit allows us to measure absorption depths in the core of the two \ion{Na}{i} D lines of $0.64\pm0.07\%$ and $0.40\pm0.07\%$ for the D2 and D1 lines, respectively.

\subsection{The \ion{Na}{i} D sodium doublet as a probe to the upper atmosphere}
We compare our measurements with transmission spectroscopy models. We use the $\eta$ model described in \citet{Ehrenreich2006,Ehrenreich2012A,Ehrenreich2014}. The model computes the opacity along the line of sight grazing the atmospheric limb of planet and integrate over the whole limb. The planetary atmosphere is modeled as a perfect gas in hydrostatic equilibrium composed by 93$\%$ of molecular hydrogen and, 7$\%$ of helium, and solar abundance of atomic sodium (volume mixing ratio of $10^{-6}$). Several model atmospheres are simulated with isothermal temperature profiles for temperature ranging from 1\,000~K to 3\,600~K by step of 100~K. To increase the resolution in temperature, we interpolate linearly between the models. The sodium line profiles are modeled with Voigt functions resulting from the convolution of a Doppler thermal profile (dominating the line cores) and a Lorentzian profile accounting for the natural and collisional broadenings in the line wings. Collisional or pressure broadening is completely dominating the line shapes far (over scales of $\sim100$~nm) from the cores \citep{Iro2005}. The half-width at half-maximum (HWHM) of the pressure broadened Lorentzian follows the prescriptions of \citet{Burrows2000} and \citet{Iro2005}, which are valid over the wavelength range we are studying here. The modeled transmission spectra, calculated at a resolution of $0.01\AA$, are convolved with the average HARPS Line Spread Function (LSF), which is well represented by a Gaussian with width of $0.05\AA$. In contrast with HST/STIS, the comparison between the models and our data is eased by the fact that HARPS resolves individual lines arising in the planet atmosphere by a factor $\sim10$ (see above).

Before comparing our transmission spectrum to models, we first fitted to our data a simple haze model. Knowing that the Rayleigh slope in our wavelength range analysis is negligible, hazes can be well represented by an absorption cut-off at a given level. This level must be compatible with our measured continuum, which corresponds to the wavelength ranges of model 0 in Table~\ref{table5}. We want also to determine if the sodium wings absorption dominates the haze absorption in the transmission spectrum. The “wings' shoulders” region is described by the wavelength ranges of model 1 in Table~\ref{table5}. Therefore, the maximum altitudes of the cloud deck which is compatible with the continuum level is given by a constant fit to the wavelength regions used to fit model atmospheres 0 and 1 to the data. This constant level gives a Bayesian information criterion (BIC) of 1\,597 in the wings region (region 1).

\begin{figure}[t!]
\centering{
\includegraphics[width=0.47\textwidth]{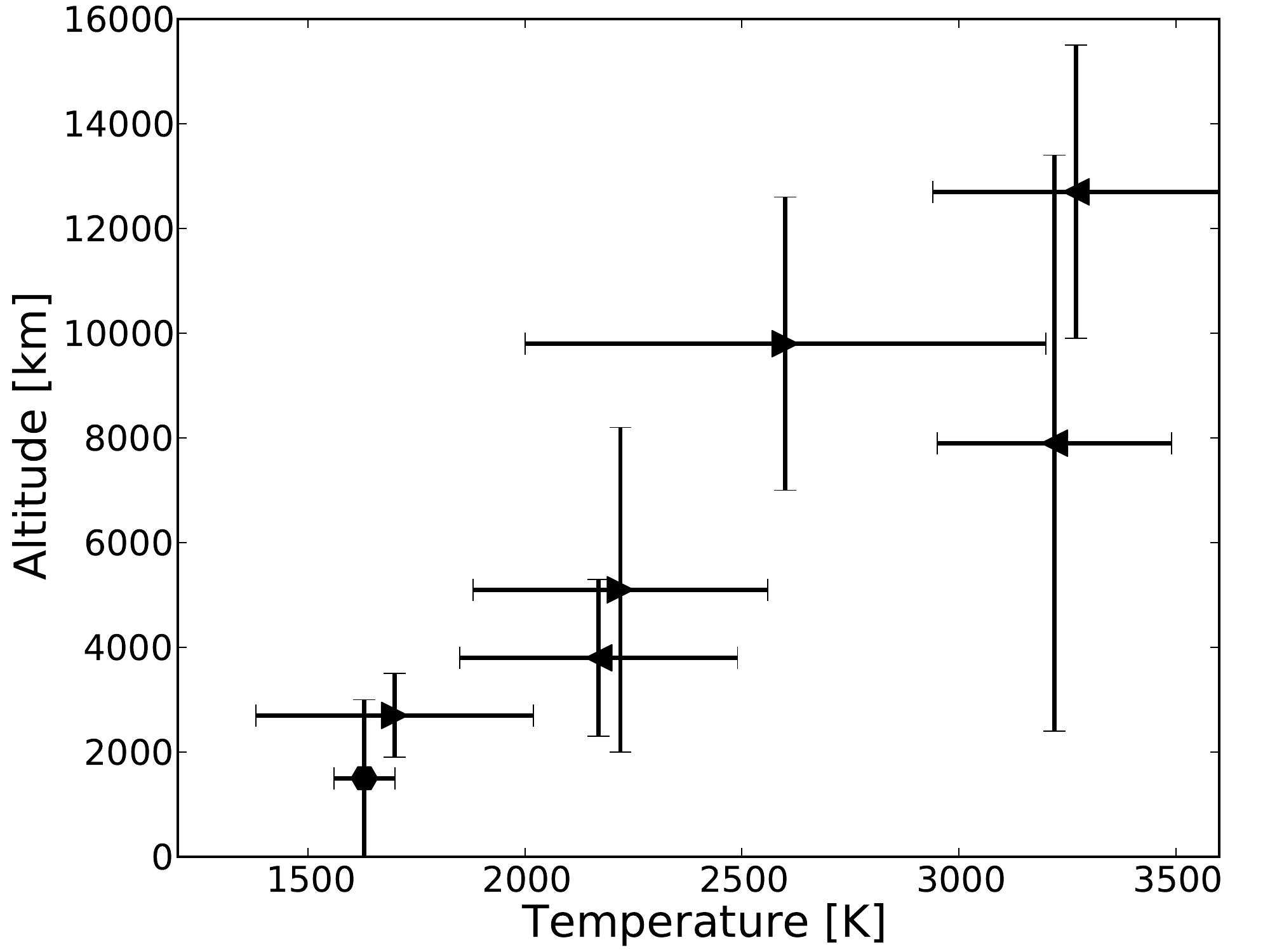}}
\caption{Temperature profile obtained by our fitting process of $\eta$ models to the transmission spectrum. The right and left triangle correspond to the \ion{Na}{i} D1 and D2 lines, respectively. The diamond correspond to the “wing'shoulders”. A temperature gradient of $\sim0.2$~K~km$^{-1}$ is measured.}
\label{Tprofile}
\end{figure}

To compare our observed transmission spectrum to models, our approach is to fit the different regions of the measured transmission spectrum around the sodium doublet with model atmospheres with different isothermal profiles. We divided the transmission spectrum in separated wavelength ranges around each line of the sodium doublet, each corresponding to altitude slices. The altitude slices are given by the maximum between the altitude errors on the data and the difference in altitudes in the given ranges. These ranges are listed in Table~\ref{table5}. We fit these different parts of the spectrum with different isothermal models, probing separately the line cores (models 4), the region encompassed within $1\times$FWHM (excluding the line cores; models 3), the region between $1\times$ and $2\times$ the FWHM (models 2), the line wings (excluding the previous regions; model 1), and the continuum (model 0). All the fits are obtained with a $\chi^2$ minimization over the grids of atmospheric models obtained by varying the temperature and a general offset in relative flux. In the continuum wavelength range we cannot see a significant sodium absorption due to absorbing wings. Then, model 0 was chosen with a fixed temperature of 1140~K (the equilibrium temperature of HD\,189733b). The fit of this model to the continuum allows us to determine the offset in planetary radius between the data and the models. We fix this offset for the subsequent model adjustments. We fit models 1, 2, 3, and 4 to their respective spectral regions, simply considering the model temperature as a free parameter. The fit of model 1 gives us a temperature of 1\,630~K for a BIC of 1\,542. The comparison with the BIC of 1\,597 obtained with a constant haze level is very strong evidence ($\Delta \mathrm{BIC}\geq 20$) in favor of a scenario where the \ion{Na}{i} wings absorption dominates the haze absorption. Thus, part of our measured sodium absorption is due to the wings and correspond to lower altitudes and temperatures. The three other models (2, 3 and 4) investigate different parts of the line cores inside 1~$\AA$ (2 FWHM, see Table~\ref{table5}). The best fit models are shown in Fig.~\ref{Fit}. For each range of altitudes, we therefore derive one temperature. The resulting temperature profile as a function of altitude is shown in Fig.~\ref{Tprofile}. The temperature linearly increases with altitude with a gradient of $\sim0.2$~K~km$^{-1}$ ($0.2\pm0.1$~K~km$^{-1}$). Note that our measured temperatures are underestimated due to the use of isothermal models, which give, for a same upper temperature, more extended atmospheres than models with positive temperature gradient.

Our high-resolution measurement of sodium in HD\,189733b, allows us to probe a new atmospheric region (between $\sim1.1-1.2$ planetary radius, corresponding to pressures of $10^{-7}-10^{-9}$~bar), above the result previously obtained at lower resolution \citep{Huitson2012}. The present study suggests that a part of the sodium absorption takes place up to the (lower) thermosphere, where heating by the stellar X/EUV photons occurs \citep{Lammer2003,Yelle2004,Vidal2011,Koskinen2013a}. Meanwhile our interpretation relies on hydrostatic models, which could be questionable at very high altitudes. Further theoretical analysis of these results is developed in a separate paper \citep[][accepted in ApJ]{Heng2015}.

\section{Conclusions}\label{Sec_Conclu}

We have presented an analysis of all the available transit observations of HD\,189733b with HARPS on ESO~3.6~m telescope. We carefully corrected for the change in radial velocity of the planet and for telluric contamination, fully exploiting the high spectral resolution and the stability of HARPS. We detect excess of absorption due to \ion{Na}{i} D lines in our transmission spectrum.  We measure line contrasts of $0.64\pm0.07\%$ (D2) and $0.40\pm0.07\%$ (D1) and FWHMs of $0.52\pm0.08~\AA$. Sodium is clearly detected over several different passbands and the signatures are robust against different statistical tests. Moreover, these detections are consistent and comparable in terms of precision with those obtained from space-borne or 10~m class ground-based facilities.

High-resolution permits us to measure a blueshift in the line positions of $0.16\pm0.04\ \AA$. We interpret it as wind in the upper layer of the atmosphere of a velocity of $8\pm2$ km/s. Since we resolved the sodium lines, we used it as a probe of the lower pressure part of the atmosphere. We measured a non-isothermal temperature profile inside our range of probed altitudes, where the sodium absorption is peaking out above the haze present in the atmosphere. We found that the temperature increases towards the thermosphere. A temperature gradient of $\sim0.2$~K~km$^{-1}$ is measured. The existence of a positive temperature gradient does not depend on assumptions about the haze layer. In fact, the gradient can be measured from the resolved line cores and in particular from the difference in absorption levels between the D1 and the D2 lines. This shows that beyond the detections of atomic species, transit spectroscopy allows characterization of physical conditions present in atmospheres.

These transit spectroscopy data were already used by \citet{Triaud2009} and by \citet{Collier2010} to study the Rossiter-McLaughlin effect. Furthermore, with these same observations, \citet[][to be submitted]{DiGloria2015} detected a slope in the planet-star radius ratios that \citet{Pont2008,Pont2013} interpreted as Rayleigh scattering. These studies demonstrate the relevance of studying exoplanet atmospheres with high-resolution spectrographs mounted on 4-meter-class telescopes. This is especially important considering that several such facilities will be built in the coming years to prepare for the follow-up of exoplanet candidates from incoming space missions such as K2 (on-going), TESS, CHEOPS, and PLATO.

In the near future, the HARPS-like instrument ESPRESSO, will be mounted on the Very-Large Telescope (VLT) incoherent focus. ESPRESSO could be used to probe exoplanetary atmospheres in the optical. This will bring a more complete census of the chemical composition and study of aeronomic properties of a large variety of exoplanets, including those with larger bulk densities or with fainter host stars. On the infrared side, several spectrographs will be built soon such as SPIROU, CARMENES, IRD and CRIRES+. Based on recent results \citep[\textit{e.g.}][]{Birkby2013,Snellen2014} a strong development of atmospheric detections and characterization in the infrared is foreseeable. Eventually, high-resolution spectra of exoplanets with broad spectral coverage will allow inventory of the atmospheric chemical composition, the exploration of aeronomic properties such as pressure-temperature profile and lead to a better understanding of the formation and evolution of exoplanets. This is also a motivation for building high-resolution spectrograph on the European Extremely-Large-Telescope \textit{e.g.} HiReS and METIS \citep{Udry2014,Brandl2014,Snellen2015}. In the meantime, extensive studies have to be done with existing facilities and data especially in the optical domain \citep[\textit{e.g.}][]{Hoeijmakers2014} to help observers make the best use of future facilities. The present study is a significant step in this direction. It could be extended to the search of other species or as a systematic search in other hot-jupiter atmospheres.

\begin{acknowledgements}
This work has been carried out within the frame of the National Centre for Competence in Research ‘PlanetS’ supported by the Swiss National Science Foundation (SNSF). The authors acknowledge the financial support of the SNSF. We thank M. Mayor, A.H.M.J. Triaud and A. Lecavelier for obtaining the data. We thank K. Heng, S. Khalafinejad, E. Di Gloria, V. Bourrier and R. Allart for discussion and insight. We gratefully acknowledge our referee, I.A.G. Snellen, for valuable comments that improved our manuscript.
\end{acknowledgements}


\bibliographystyle{aa}

\end{document}